\documentclass[a4paper,11pt]{article}
\usepackage{jheppub} 
\usepackage{graphicx, graphics, color}
\usepackage[T1]{fontenc} 
\usepackage{amsmath, amssymb, bm, graphicx, graphics, color, mathrsfs}
\usepackage{epsfig}
\usepackage{dcolumn}
\usepackage{bm}
\usepackage{tikz}
\usepackage{graphicx, graphics, color}
\usepackage{dcolumn}
\usepackage{bm}
\usepackage{hyperref}
\usepackage[mathlines]{lineno}

\newcommand{\be}{\begin{equation}}
\newcommand{\ee}{\end{equation}}
\newcommand{\ben}{\begin{eqnarray}}
\newcommand{\een}{\end{eqnarray}}

\newcommand{\cO}{{\cal O}}

\newcommand{\p}{\partial}
\newcommand{\na}{\nabla}

\newcommand{\tA}{\tilde A}

\newcommand{\ep}{\epsilon}

\newcommand{\ga}{\gamma}

\newcommand{\tD}{{\tilde D}}
\newcommand{\tB}{{\tilde B}}
\newcommand{\tE}{{\tilde E}}
\newcommand{\talpha}{{\tilde \alpha}}

\newcommand{\tga}{{\tilde \gamma}}

\newcommand{\tc}{\tilde c}
\newcommand{\td}{\tilde d}

\newcommand{\zz}{\mathbb{Z}_2}

\title{
Magnetotransport of Weyl semimetals with $\mathbb{Z}_2$ topological charge and chiral anomaly.}

\author[1]{Marek Rogatko\note{rogat@kft.umcs.lublin.pl, marek.rogatko@poczta.umcs.lublin.pl}}
\author[2]{Karol I. Wysokinski\note{karol@tytan.umcs.lublin.pl}}
\affiliation{Institute of Physics \\
Maria Curie-Sk{\l}odowska University \\
20-031 Lublin, pl. Marii Curie-Sk{\l}odowskiej 1, Poland}

\abstract{
We calculate  the magnetoconductivity of the
Weyl semimetal with $\mathbb{Z}_2$ topological charge and chiral anomaly utilizing the recently
developed hydrodynamic theory.  The system in question will be influenced by magnetic fields 
connected with ordinary Maxwell and the second $U(1)$-gauge field,
which couples to the anomalous topological charge.
The presence of chiral anomaly and $\mathbb{Z}_2$ topological charge endow
the system with new transport coefficients. We start with the linear perturbations of the
hydrodynamic equations and calculate the magnetoconductivity of this system.
The holographic approach in the probe limit is implemented to obtain the explicit dependence of the longitudinal magnetoconductivities on the magnetic fields. 
 }

\keywords{Gauge-gravity correspondence,
Holography and condensed matter physics (AdS/CMT), Black Holes}

\begin{document} 

\maketitle
\flushbottom



\section{Introduction}
\label{sec:intro}

The recently observed resurgence of interest in the physics of chiral systems 
is related to the important discoveries in such fields as dynamics of relativistic
quark-gluon plasma \cite{kha06}-\cite{kha14} and  high energy astrophysical processes in proto-neutron stars \cite{kam15}
on one side and
the recent realization of the known and novel relativistic symmetries for electrons in condensed matter
systems. 
Chiral magnetic and vortical effects observed in the high energy nuclear
collisions have been reviewed recently \cite{kharzeev2016}. The topological  quantum matter is becoming 
an important play-ground for discoveries of numerous novel 
phenomena not observed before in particle physics.   

Electrons in metals form a system of interacting particles.  The strength of the interactions
is  measured with respect to their kinetic energy and thus depends on the properties of the system under study. 
The standard paradigm was to describe electrons in metals by the Landau quasiparticle theory.
However, there are many very strongly interacting materials which descriptions require approaches going beyond quasiparticle picture.   
For example, in very clean metals the flow of electrons  resembles that of the classical fluid
and is described by the laws of classic hydrodynamic theory \cite{mol2016}. 
 In some materials like Dirac or Weyl semimetals \cite{wehling2014}, the spectrum of electrons in the vicinity of the Fermi level,
is linear and thus relativistic despite their velocity $v_F$ being much smaller then the
velocity of light.  Thus, the condensed matter allows for the relativistic
symmetries and opens new possibilities for laboratory studies of exotic particles, which were predicted, 
but never seen in vacuum.  The long sought Majorana fermion may constitute an example of them \cite{elliott2015}.


The discovery of the aforementioned semimetals of Dirac or Weyl character, has sparked the intense studies of band crossing phenomena
in crystals. This resulted in the discovery of ``relativistic'' quasiparticles with effective ``spin''
quantum number different from $1/2$. The crystal space groups hosting such fermions have been identified and
the materials realizing them proposed \cite{bradlyn2016}. Moreover, the ``spin'' 3/2 electrons were apparently
 observed \cite{kim2018}. 

The relativistic spectrum together with the associated chirality of the quantum particles is a source of new
phenomena generally known as chiral quantum anomalies, first discovered by   Adler \cite{adler1969} 
and independently by  Bell  and  Jackiw \cite{bell1969}.  They
have played an important role in understanding the neutral pion decay and recently they are subject of 
vigorous studies in condensed matter laboratories. In the presence of dense matter, quantum anomalies modify
the relativistic hydrodynamics by leading to new phenomena and novel transport coefficients \cite{son09}, i.e., an anomalous current is 
generated by an external magnetic field or by vortices in fluid which carries 
the charge in question \cite{ama11}.

In the realm of condensed matter physics the role of chiral and other anomalies is especially visible
under the influence  of external magnetic field or  a longitudinal temperature gradient. 
The anomaly also influences the electric DC conductivity. Namely, the longitudinal DC conductivity 
is amplified by magnetic field \cite{nie83,son13}. 
The key prediction was found in a kinetic description at weak coupling, as well as, hydrodynamics and
holographic attitude at strong coupling.

The problem of the longitudinal magnetoconductivity, in the system with a chiral anomaly
and background magnetic field was examined in \cite{lan15}. By means of the linear response
method in the hydrodynamic limit and holography attitude, it was shown that one needs to have
energy, momentum and charge dissipations to obtain a finite DC longitudinal magnetoconductivity.
The same problem was studied in \cite{cho15}, where the longitudinal DC conductivity bounded with the Lifshitz like fixed points, in the presence of
chiral anomalies, in $(3+1)$-dimensions was found.

Following the method presented in 
\cite{don14}, the generalized expression for DC and Hall conductivities were delivered also for large class
of the holographic massive models \cite{zho15}.
Massive gravity models \cite{bla13}-\cite{luc14}
attract attentions due to the diffeomorphism symmetry breaking
which causes the non-conservation of the energy-momentum tensor. The elaborated effect is similar
to the dissipation of the momentum.
  
Among all, the holographic treatment of the system in question reveals that in the intermediate 
regime of the magnetic field, one gets a negative magnetoresistivity decreasing as an inverse 
of the magnetic field.	
	The specific angular dependence of this enhancement, in the longitudinal direction along
the magnetic field, is the key prediction which arises from the aforementioned descriptions.
However, already in the weak coupling limit, it was envisaged that negative magnetoresistance
might appear \cite{gos15}.

Similar effect can arise in the hydrodynamical attitude, when the distinctive gradient expansion is taken
into account (the non-relativistic constitutive relations can be derived from the most general covariant
form of the expression, up to the third order in the field strength) \cite{bau17}. Moreover,
in a relativistic hydrodynamic theory of transport, it depends on the macroscopic model (one can achieve both negative and positive magnetoresistance). In non-Galilean
invariant fluids the effect can arise not only due to the presence of a background magnetic field but also depends on the specific structure of the considered 
hydrodynamics \cite{bau17}.

The other mechanism leading to the negative magnetoresistivity was 
proposed in \cite{rog18}-\cite{rog18a}, by using two interacting $U(1)$-gauge fields.

The recent experimental works conducted in Dirac or Weyl semimetals like $Na_3Bi$,~$Zr Te_5$ \cite{xio15,li16}
and $Ta As$,~$Nb P$  \cite{zha15}-\cite{goo17}, confirm the evidences of chiral anomaly. In principle
there are two classes of Dirac semimetals, in first one the Dirac points appear
at the time reversal invariant
momenta in the first Brillouin zone. The second class comprises the elements in which  
the Dirac points take place in pairs at two arbitrary points in the Brillouin zone. 
They are separated in the momentum 
space along a rotational axis \cite{young2012,chiu2016}. Moreover it turns out that they are
characterized by a non-trivial $\zz$ topological invariant 
protecting the nodes and leading to the presence of Fermi arc surface states \cite{yan14}-\cite{kob15}. 
It has been argued that the $\zz$ anomalous charge affects the transport characteristics
of the materials in question \cite{burkov2016}.
The $\zz$ topological charge influence on the transport properties was studied
in a relativistic hydrodynamics limit in \cite{rog18b}.

The behavior of metals influenced by electromagnetic field is of the great importance 
for understanding their transport properties. For the most materials under inspection the 
longitudinal conductivity is a decreasing function of the magnetic field. However, for the Weyl 
semimetals (materials for which conduction bands intersect at distant 
points in the Brillouin zone) one has an exception to the rules \cite{lan15}.
In the recent experiments \cite{kim13} the negative magnetoresistance
in the intermediate regime of the magnetic field was confirmed. There were proposed several alternative ways
of explaining the above phenomenon. For example in \cite{kim13}-\cite{kim14}, the weak-antilocalization
effect was considered for the explanation of the aforementioned effect.

The main objectives of our work are to examine the influence of $\zz$ topological charge
and chiral anomaly on conductivities, in the presence of non-zero ingredients of magnetic field
components of the $U(1)$-gauge fields. These considerations constitute
the development of the ideas \cite{rog18b}, where the hydrodynamical characteristics of the model
with chiral anomaly and $\zz$ topological charge were presented.


The paper is organized as follows.
In Section \ref{sec:model} we present the calculations 
of electric conductivities for both studied $U(1)$-gauge fields, in the hydrodynamics limit,
taking into account chiral anomaly and $\zz$ topological charge. We have elaborated the case with background
magnetic fields and with all the possible dissipation terms in order to achieve finite 
value of the DC longitudinal conductivities. Section 3 is devoted to the holographic 
model of our
 system, in the probe limit. We discuss the relation  between the parameters entering 
the holographic action and those responsible for anomalies in section 3.1.
As a background spacetime we take five-dimensional AdS Schwarzschild black brane.
Section 4 is connected with the independent calculation of magnetoconductivities 
for the studied holographic model with chiral anomaly and $\zz$ topological charge
in the probe limit, using Kubo formula. 
The obtained results match  the hydrodynamical formula in the appropriate limit.
In last section we conclude the main results.

\section{Hydrodynamics of the system}
\label{sec:model}
The presence of quantum  triangle anomalies severely modifies the relativistic hydrodynamic equations.
The novel transport coefficients appear and are expressed in terms of anomaly coefficients (charges) and the
systems equation of state \cite{son09}. Recently, we have generalized  the relativistic hydrodynamic theory to 
 the systems endowed with chiral and topological charges  \cite{burkov2016}. To describe $\zz$ topological charge 
in chiral Weyl semimetal of the second kind we have used two different electro-magnetic fields.

The motivation standing behind our research is to include the different dissipation 
terms in studies of longitudinal DC-conductivity. At first we directly find
the aforementioned conductivity in the hydrodynamical attitude in the four-dimensional 
spacetime with two background magnetic fields. One of the fields is connected with the ordinary Maxwell 
one while the other constitutes the magnetic component of the auxiliary $U(1)$-gauge 
field which couples to the anomalous $\zz$ topological charge. 
We restrict our considerations to the linear response level.
For the purpose of the future reference and to established the notation we shall briefly present 
main ideas and results of the mentioned hydrodynamic approach \cite{rog18b}.

The inclusion of the dissipation terms \cite{har07,lan15}, is to perturb first a hydrodynamical 
system in a given equilibrium state and solve the equation of motion with initial values of the perturbations.
In the next step, the searched transport coefficients can be found using the response of the electric 
thermal currents  to the initial values of the corresponding perturbations.
\subsection{Hydrodynamic equations for system with chiral and $\zz$ topological charge}
\label{sec:hydreq}
The key set of the studied relations will be the hydrodynamical equations of motion 
in the presence of $\mathbb{Z}_2$ topological charge and chiral anomaly, provided by \cite{rog18b}
\ben \label{jt1}
\p_\alpha T^{\alpha \beta}(F,B) &=& F^{\beta \alpha} j_{\alpha}(F) + B^{\beta \alpha} j_{\alpha}(B),\\ \label{jf1}
\p_\alpha j^\alpha (F) &=& C_1~E_{(F)\alpha} B^{(F) \alpha} + C_2~\tE_{(B)\alpha} \tB^{(B)\alpha},\\ \label{jb1}
\p_\alpha j^\alpha (B) &=& C_3~\tE_{(B)\alpha} B^{(F)\alpha} + C_4~E_{(F) \alpha} \tB^{(B)\alpha}.
\een
It has to be recalled that $C_2=C_4$ due to symmetry reasons.
In the above equations we have denoted the electric and magnetic fields in the fluid rest frame by
\ben
E^{(F)}_{\alpha} &=& F_{\alpha \beta} u^\beta, \qquad B^{(F)}_{\alpha} = \frac{1}{2} \ep_{\alpha \beta \rho \delta}~u^\beta~F^{\rho \delta}, \\
\tE^{(B)}_{\alpha} &=& B_{\alpha \beta} u^\beta    \qquad  \tB^{(B)}_{\alpha} = \frac{1}{2} \ep_{\alpha \beta \rho \delta}~u^\beta~B^{\rho \delta}.
\een
 where $F_{\mu \nu} = 2 \p_{[ \mu} A_{\nu ]}$ stands for the ordinary Maxwell field strength tensor, while
the second $U(1)$-gauge field $B_{\mu \nu}$ is given by $B_{\mu \nu} = 2 \p_{[ \mu} B_{\nu ]}$.
$j_\mu(F),~j_\mu (B)$ represent the adequate currents connected with each of the gauge field.

On the other hand, the energy momentum tensor and the adequate currents needed for the hydrodynamic 
description of the relativistic fluid, imply \cite{landau1959}
\ben
T^{\mu \nu} &=& \Big( \ep + p \Big) u^\mu u^\nu + p~ g^{\mu \nu} + \tau^{\mu \nu},\label{tmn}\\
j^\mu (F) &=& \rho~u^\mu + V_F^\mu,\label{jf}\\
j^\mu (B) &=& \rho_d~u^\mu + V_B^\mu,
\label{jb}
\een
where $\tau^{\mu \nu}$ and $V_{F(B)}^\mu$ denote corrections, higher order  in velocities, responsible for dissipative effects.
Using the fact that there is no dissipative force in the rest frame of the liquid element, we obtain
$u_\alpha~\tau^{\alpha \beta }=  u_\alpha~V_F^\alpha =u_\alpha~V_B^\alpha   = 0$.
By $\ep$ we have  denoted the energy density, $p$ is connected with pressure and $\rho$,~$\rho_d$ are the $U(1)$ 
charge densities.  The four-vector $u^\mu$, with the 
normalization $u_\mu u^\mu =-1$, describes the flow of the fluid in the system in question and the general expressions 
for the dissipative components of the energy - momentum tensor and currents read
\ben
\tau^{\mu \nu} &=& - \eta P^{\mu \alpha} P^{\nu \beta} \Big( \p_\alpha u_\beta + \p_\beta u_\alpha \Big) - \Big( \zeta - \frac{2}{3} \eta \Big) P^{\mu \nu} \p_\alpha u^\alpha,\\ \nonumber
V_F^\alpha &=& - \sigma_F~\bigg[ T~P^{\alpha \beta}~\p_\beta \Big( \frac{\mu}{T} \Big) - E^{(F)\alpha} \bigg] 
- \sigma_{F \tB}~\bigg[ T~P^{\alpha \beta}~\p_\beta \Big( \frac{\mu_d}{T} \Big) - \tE^{(B)\alpha} \bigg]  + \xi~\omega^\alpha  \\ 
&+& \xi_B~B^{(F)\alpha} + \xi_{F \tB} \tB^{(B)\alpha},\\ \nonumber \label{vb}
V_B^\alpha &=& - \sigma_B~\bigg[ T~P^{\alpha \beta}~\p_\beta \Big( \frac{\mu_d}{T} \Big)  -\tE^{(B)\alpha} \bigg] 
- \sigma_{B F}~\bigg[ T~P^{\alpha \beta}~\p_\beta \Big( \frac{\mu}{T} \Big)  - E^{(F)\alpha} \bigg] + \xi_d~\omega^\alpha \\ 
&+& \xi_\tB~\tB^{(B)\alpha} + \xi_{\tB F} B^{(F)\alpha}.
\een
where $P^{\mu \nu} = g^{\mu \nu} + u^\mu u^\nu$,  by $\omega_\mu = 1/2 \ep_{\mu \nu \rho \delta}  u^\nu  \p^\rho u^\delta$ we have denoted the vorticity.
On the other hand, $\xi,~\xi_d,~\xi_B,~\xi_\tB,~\xi_{F \tB},~\xi_{\tB F}$ are kinetic coefficients being functions of $T$ and $\mu,~\mu_d$. They are given by the following expressions
 \cite{rog18b}:
\ben \label{eq267} \nonumber
\xi&=&C_1\mu^2\bigg(1-\frac{2}{3}\frac{\rho ~\mu}{\ep+p}\bigg) + \mu_d^2 C_2 \bigg( 1 - 2 \frac{\rho ~\mu}{\ep + p} \bigg)
- 4 \frac{\rho~T^2}{\ep + p} \bigg( \mu \tga_1 + \mu_d \tga_2 \bigg) \\
&-&  2 \frac{\rho ~T^3}{\ep + p} \tga_3 + 2 \tga_1 ~T^2, \\
\label{eq268} 
\xi_d&=& -\frac{2}{3}C_1\frac{\rho_d~ \mu^3}{\ep+p} + 2 C_2~ \mu~\mu_d \bigg(1-\frac{\rho_d~\mu_d}{\ep+p}\bigg) 
- 4 \frac{\rho_d~T^2}{\ep + p} \bigg( \mu \tga_1 + \mu_d \tga_2 \bigg) \\ \nonumber
&-&  2 \frac{\rho_d ~T^3}{\ep + p} \tga_3 + 2 \tga_2~ T^2,
\\
\label{eq269}
\xi_B&=& C_1 \mu  \bigg(1-\frac{1}{2}\frac{\rho~ \mu}{\ep+p}\bigg) - \frac{1}{2}C_3\frac{\rho ~\mu_d^2}{\ep+p} - \frac{\rho ~T^2}{\ep + p} \tga_1,\\
\label{eq270}
\xi_\tB&=& C_2 \mu \bigg(1-\frac{\rho_d ~\mu_d}{\ep+p}\bigg) - \frac{\rho_d~ T^2}{\ep + p} \tga_2.\\
\label{eq270a}
\xi_{F \tB} &=& C_2 \mu_d \bigg( 1 - \frac{\rho~ \mu}{\ep + p} \bigg) - \frac{\rho~T^2}{\ep + p} \tga_2,\\
\xi_{\tB F} &=& C_3 \mu_d \bigg( 1 - \frac{1}{2} \frac{\rho_d ~\mu_d}{\ep + p} \bigg) - \frac{1}{2} C_1 \frac{\rho_d ~\mu^2 }{\ep+p} 
- \frac{\rho_d~T^2}{\ep +p}~\tga_1.
\label{eq270b}
\een
Let us discuss the hydrodynamical regime of the theory in  question. Having in mind the relations for $V_F^\alpha $ and $V_B^\alpha $, suppose that
$T \gg \mu,~T\gg \mu_d$, as well as, $E^{(F)\alpha}, ~\tE^{(B)\alpha},~B^{(F)\alpha},~\tB^{(B)\alpha}  \ll T^2$. These requirements help us to assure that the
first derivative expansions constitute the leading order contributions in $V_F^\alpha $ and $V_B^\alpha $. 

As in \cite{lan15}, one also assumes that $\mid \tc~ B^{(F)\alpha} \mid \ll T^2$ and $\mid \td~ \tB^{(F)\alpha} \mid \ll T^2$, due to the fact that
in the first derivative expansions in $j^\mu (F)$ and $j^\mu (B)$, the magnetic components $B^{(F)\alpha},~\tB^{(B)\alpha}$ enter as the contributions of 
the forms $\tc ~B^{(F)\alpha},~\td~ \tB^{(F)\alpha} $, respectively. By $\tc,~\td$ we have denoted dimensionless numbers, related to the charges $C_i$ which can be either of large or small values.

In our considerations we shall suppose that the system in question
is in an equilibrium state in the grand canonical ensemble with chemical potentials $\mu,~\mu_d$, temperature $T$ and the local velocity will fulfill the condition
$u^t =1$. We also assume that 
\be
\ep + p = T s + \mu ~\rho + \mu_d~\rho_d, \qquad d p = s dT + \rho d \mu + \rho_d d\mu_d.
\ee

\subsection{Hydrodynamic magnetotransport of the strongly interacting system}

In order to find the magnetic field dependence of the electrical conductivity 
in the system with chiral anomaly and $\zz$ topological charge, we suppose that one deals with a background magnetic fields 
 in $z$-direction, provided by 
\be
A_2 = B~x, \qquad B_2 = \tB_{add}~x,
\ee
and consider the case when $E^{(F)\mu} = F^{t \mu} = 0$ and $\tE^{(B)\mu} = B^{t \mu} = 0$.
We shall study the response of the currents to the perturbations of $\delta E^{(F)z}$ 
and $\delta \tE^{(B) z}$ fields.

In the thermodynamical system under consideration the perturbations of the thermodynamical 
variables are given by
\ben
\mu(x_\alpha) &=& \mu + \delta \mu(x_\alpha), \qquad \mu_d(x_\alpha) = \mu_d + \delta \mu_d(x_\alpha),\nonumber \\
T(x_\alpha) &=& T + \delta T(x_\alpha), \qquad u^\beta (x_\alpha) = \Big( 1,~\delta u_j (x_\alpha) \Big).
\een
Moreover, the perturbations of the other components of the gauge fields are chosen as follows:
\ben
\delta E^{(F)z} &=& \delta F^{tz}, \qquad \delta \tE^{(B) z} = \delta B^{tz},\label{pert1} \\
\delta E^{(F) x} &=& \delta F^{tx} + B~\delta u^y, \qquad \delta \tE^{(B) x} = \delta B^{tx} + \tB_{add}~\delta u^y, \label{pert2}\\
\delta E^{(F) y} &=& \delta F^{ty} - B~\delta u^x, \qquad \delta \tE^{(B) x} = \delta B^{ty} - \tB_{add}~\delta u^x.
\label{pert3}
\een
Having in mind the above forms of perturbations of the hydrodynamical variables, to the linear order, 
the perturbations of the conserved quantities fulfill
the following relations:
\ben
\delta T^{00} &=& \delta \ep,\\
\delta T^{0i} &=& \Big( \ep + p \Big) \delta u^i,\\
\delta T^{ij} &=& \delta p~g^{ij} - \eta \Big( \p^i \delta u^j + \p^j \delta u^i - \frac{2}{3} g^{ij} \p_m \delta u^m \Big) - \zeta g^{ij} \p_m \delta u^m,
\een
and the modifications of the respective currents connected with $U(1)$-gauge fields are calculated by combining
equations (\ref{pert1})-(\ref{pert3}), having in mind the relations (\ref{tmn})-(\ref{jb}). Consequently we obtain  
\ben
\delta j^0(F) &=& \delta \rho + \xi_B~\delta u_z~B + \xi_{F \tB}~\delta u_z~\tB_{add},\\ \nonumber
\delta j^x (F) &=& \rho \delta u^x - \sigma_F T \p^x \Big( \delta \big( \frac{\mu}{T} \big) \Big) +\sigma_F \Big( \delta F^{tx} + B \delta u^y \Big) 
- \sigma_{F \tB} T \p^x \Big( \delta \big( \frac{\mu_d}{T} \big) \Big) \\
&+& \sigma_{F \tB} \Big( \delta B^{tx} + \tB_{add} \delta u^y \Big) 
+ \xi~\p_{[y } \delta u_{z]},\\ \nonumber
\delta j^y (F) &=& \rho \delta u^y - \sigma_F T \p^y \Big( \delta \big( \frac{\mu}{T} \big) \Big) +\sigma_F \Big( \delta F^{ty} - B \delta u^x \Big) 
- \sigma_{F \tB} T \p^y \Big( \delta \big( \frac{\mu_d}{T} \big) \Big) \\
&+& \sigma_{FB} \Big( \delta B^{ty} - \tB_{add} \delta u^x \Big) 
+ \xi ~\p_{[z } \delta u_{x]},\\ \nonumber
\delta j^z(F) &=& \rho \delta u^z - \sigma_F T \p^z \Big( \delta \big( \frac{\mu}{T} \big) \Big) +\sigma_F \delta F^{tz}  
- \sigma_{F \tB} T \p^z \Big( \delta \big( \frac{\mu_d}{T} \big) \Big) \\
&+& \sigma_{F \tB} \delta B^{tz} + B \delta \xi_B + \tB_{add} \delta_{F \tB} + \xi ~\p_{[y } \delta u_{z]}.
\een
The current perturbations of the additional gauge field imply
\ben
\delta j^0(B) &=& \delta \rho_d + \xi_\tB~\delta u_z~\tB_{add} + \xi_{\tB F} ~\delta u_z~B,\\ \nonumber
\delta j^x (B) &=& \rho_d \delta u^x - \sigma_B T \p^x \Big( \delta \big( \frac{\mu_d}{T} \big) \Big) +\sigma_B \Big( \delta B^{tx} + \tB_{add} \delta u^y \Big) 
- \sigma_{B F} T \p^x \Big( \delta \big( \frac{\mu}{T} \big) \Big) \\
&+& \sigma_{BF} \Big( \delta F^{tx} + B \delta u^y \Big) 
+ \xi_d ~\p_{[y } \delta u_{z]},\\ \nonumber
\delta j^y (B) &=& \rho_d \delta u^y - \sigma_B T \p^y \Big( \delta \big( \frac{\mu_d}{T} \big) \Big) +\sigma_B \Big( \delta B^{ty} - \tB_{add} \delta u^x \Big) 
- \sigma_{B F} T \p^y \Big( \delta \big( \frac{\mu}{T} \big) \Big) \\
&+& \sigma_{B F} \Big( \delta F^{ty} - B \delta u^x \Big) 
+ \xi_d ~\p_{[z } \delta u_{x]},\\ \nonumber
\delta j^z(B) &=& \rho_d \delta u^z - \sigma_B T \p^z \Big( \delta \big( \frac{\mu_d}{T} \big) \Big) +\sigma_B \delta B^{tz} 
- \sigma_{B F} T \p^z \Big( \delta \big( \frac{\mu}{T} \big) \Big) \\ 
&+& \sigma_{B F} \delta F^{tz} + \tB_{add} \delta \xi_{\tB} + B \delta \xi_{\tB F} + \xi_d ~ \p_{[y } \delta u_{z]},
\een
where, as in \cite{lan15}, we suppose that $\delta \ep,~\delta p$ are entirely impelled  
by $\delta \mu,~\delta \mu_d,~\delta T$. Moreover, one neglects the chiral vortical effects because 
they do not influence the results.

The time evolution of the perturbations is obtained from the conservation equations (\ref{jt1})-(\ref{jb1}).
To calculate them, in analogy to \cite{lan15}, we introduce the dissipation terms into the perturbed 
conservation laws of the considered theory and obtain
\ben
\p_\mu \delta T^{\mu 0} &=& \delta F^{0z} j_z(F) + \delta B^{0z} j_z(B) + \frac{1}{\tau_e} \delta T^{\ga 0} u_\ga,\\
\p_\mu \delta T^{\mu k} &=& \rho \delta F^{0 k} + \rho_d \delta B^{0 k} + F^{k \alpha} \delta j_\alpha (F) +  B^{k \alpha} \delta j_\alpha (B)
+ \frac{1}{\tau_m} \delta T^{\ga k} u_\ga,\\
\p_\mu \delta j^\mu (F) &=& C_1~\delta E_{(F)\mu}~ B^{(F)\mu} + C_2~\delta \tE_{(B)\mu}~ \tB^{(B)\mu} + \frac{1}{\tau_c} \delta j^\ga (F) u_\ga, \\
\p_\mu \delta j^\mu (B) &=& C_3~\delta \tE_{(B)\mu}~ B^{(F)\mu} + C_4~\delta E_{(F)\mu}~ \tB^{(B)\mu} + \frac{1}{\tau_{c_d}} \delta j^\ga (B) u_\ga,
\een
where we have denoted by $\tau_e$ the energy relaxation time, $\tau_m$ stands for the momentum relaxation time, while $\tau_c$ and ${\tau_{c_d}}$ describe the charge relaxation times
for the Maxwell  and the additional $U(1)$-gauge field. One should have in mind that the relaxation terms affect only the deviations from the equilibrium. 

In the next step let us put the perturbations into the energy-momentum and currents perturbation relations, in order to achieve the set of relations for $\delta \mu,~\delta \mu_d,~\delta T $ and 
$\delta u^\alpha$. For the perturbations of the energy-momentum tensor one obtains
\ben \label{eqe} \nonumber
\Big( \p_t &+& \frac{1}{\tau_e} \Big) \delta \ep + \p_i \Big[ (\ep + p) \delta u^i \Big]  - \delta E^{(F) z} \Big(  \xi_B~ B + \xi_{F \tB} \tB_{add} \Big) \\
&-& \delta \tE^{(B) z} \Big(  \xi_\tB~ \tB_{add} + \xi_{\tB F} B \Big) = 0,\\ \label{eqx}
\Big( \p_t &+& \frac{1}{\tau_m} \Big) (\ep + p) \delta u^x - \rho \delta F^{0x} - \rho_d \delta B^{0x} + \p^x \delta p 
- \eta \Big[ \p_i \p^i \delta u^x + \frac{1}{3}\p^x \p_m \delta u^m \Big] \\ \nonumber
&-& \zeta \p^x \p_m \delta u^m = 
B \Big[ \rho \delta u_y - \sigma_F T \p_y\Big( \delta \big( \frac{\mu}{T} \big) \Big) + \sigma_F \Big( \delta F^{0y} - B \delta u^x \Big) \\ \nonumber
&-& \sigma_{F \tB} T \p_y\Big( \delta \big( \frac{\mu_d}{T} \big) \Big) + \sigma_{F \tB} \Big( \delta B^{0y} - \tB_{add} \delta u^x \Big)
+ \xi \p_{[z} \delta u^{x]} \Big] \\ \nonumber
&+& \tB_{add} \Big[ \rho_d \delta u_y - \sigma_B T \p_y\Big( \delta \big( \frac{\mu_d}{T} \big) \Big) + \sigma_B \Big( \delta B^{0y} - \tB_{add} \delta u^x \Big) \\ \nonumber
&-& \sigma_{BF} T \p_y\Big( \delta \big( \frac{\mu}{T} \big) \Big) + \sigma_{BF} \Big( \delta F^{0y} - B \delta u^x \Big)
+ \xi_d \p_{[z} \delta u^{x]} \Big],\\ \label{eqy}
\Big( \p_t &+& \frac{1}{\tau_m} \Big) (\ep + p) \delta u^y - \rho \delta F^{0y} - \rho_d \delta B^{0y} + \p^y \delta p 
- \eta \Big[ \p_i \p^i \delta u^y + \frac{1}{3}\p^y \p_m \delta u^m \Big] \\ \nonumber
&-& \zeta \p^y \p_m \delta u^m = 
- B \Big[ \rho \delta u_x - \sigma_F T \p_x \Big( \delta \big( \frac{\mu}{T} \big) \Big) + \sigma_F \Big( \delta F^{0x} + B \delta u^y \Big) \\ \nonumber
&-& \sigma_{F \tB} T \p_x \Big( \delta \big( \frac{\mu_d}{T} \big) \Big) + \sigma_{F \tB} \Big( \delta B^{0x} + \tB_{add} \delta u^y \Big) 
+ \xi \p^{[y} \delta u_{z]} \Big] \\ \nonumber
&-& \tB_{add} \Big[ \rho_d \delta u_x - \sigma_B T \p_x \Big( \delta \big( \frac{\mu_d}{T} \big) \Big) + \sigma_B \Big( \delta B^{0x} + \tB_{add} \delta u^y \Big)\\ \nonumber
&-& \sigma_{BF}T \p_y\Big( \delta \big( \frac{\mu}{T} \big) \Big)  + \sigma_{BF} \Big( \delta F^{0x} + B \delta u^y \Big)
+ \xi_d \p^{[y} \delta u_{z]} \Big],\\ \label{eqz}
\Big( \p_t &+& \frac{1}{\tau_m} \Big) (\ep + p) \delta u^z - \rho \delta F^{0z} - \rho_d \delta B^{0z} + \p^z \delta p 
- \eta \Big[ \p_i \p^i \delta u^z + \frac{1}{3}\p^z \p_m \delta u^m \Big] \\ \nonumber
&-& \zeta \p^z \p_m \delta u^m = 0.
\een 
Consequently, for perturbations of the currents bounded with $F_{\mu \nu}$ and $B_{\mu \nu}$ gauge fields, we arrive at the following relations:
\ben \label{c1} \nonumber
\Big( \p_t &+& \frac{1}{\tau_c} \Big) \Big( \delta \rho + \xi_B ~B ~\delta u_z + \xi_{F \tB} ~\tB_{add} ~\delta u_z \Big) + \p_i \Big[
\rho \delta u^i - \sigma_F T \p^i \Big( \delta \big( \frac{\mu}{T} \big) \Big) + \sigma_F  \delta F^{0i}  \\ 
&-& \sigma_{F \tB} T \p^i \Big( \delta \big( \frac{\mu_d}{T} \big) \Big) + \sigma_{F \tB}  \delta B^{0i} \Big]  + \p_i \Big( \frac{\xi}{2} \ep^{ijk} \p_j \delta u_k \Big) \\ \nonumber
&+&  \p_x \Big( \sigma_F~B~\delta u^y + \sigma_{F \tB}~\tB_{add}~\delta u^y \Big) 
- \p_y \Big( \sigma_F ~B~\delta u^x + \sigma_{F \tB} ~\tB_{add}~\delta u^x \Big) \\ \nonumber
&+& \p_z \Big( B~\delta \xi_B + \tB_{add}~\delta \xi_{F \tB} \Big) -
C_1~B~\delta E^{(F)}_{ z} - C_2~\tB_{add}~\delta  \tE^{(B) }_{z} = 0, \\ \label{c2} \nonumber
\Big( \p_t &+& \frac{1}{\tau_{c_{d}}} \Big) \Big( \delta \rho_d + \xi_\tB ~\tB_{add} ~\delta u_z + \xi_{\tB F} ~B ~\delta u_z \Big) 
+ \p_i \Big[
\rho_d \delta u^i - \sigma_B T \p^i \Big( \delta \big( \frac{\mu_d}{T} \big) \Big) + \sigma_B  \delta B^{0i} \\ 
&-& \sigma_{B F} T \p^i \Big( \delta \big( \frac{\mu}{T} \big) \Big) + \sigma_{B F}  \delta F^{0i} \Big]
+ \p_i \Big( \frac{\xi_d}{2} \ep^{ijk} \p_j \delta u_k \Big) 
 \\ \nonumber
&+& \p_x \Big( \sigma_B~\tB_{add}~\delta u^y + \sigma_{BF}~B~\delta u^y \Big) - \p_y \Big( \sigma_B ~\tB_{add}~\delta u^x 
+ \sigma_{BF}~B ~\delta u^x
\Big) \\ \nonumber
&+& \p_z \Big( \tB_{add}~\delta \xi_{\tB} + B~\delta \xi_{\tB F} \Big) 
- C_3~B~\delta \tE^{B)}_z - C_4~\tB_{add}~\delta  E^{(F)}_z = 0.
\een

Further,  one can readily verify that the implementation of the Laplace transformation in $t$-direction reveals
\ben \nonumber
\omega_e \delta \ep &-& i \delta \ep^{(0)} + i (\ep + p)
\p_i  \delta u^i - i \delta E^{(F) z} \Big(  \xi_B B + \xi_{F \tB} \tB_{add} \Big) \\
&-& i \delta \tE^{(B) z} \Big( \xi_\tB  \tB_{add} 
+ \xi_{\tB F} B \Big) = 0,\\ \nonumber
(\ep &+& p) \Big[ \omega_m  \delta u^x - i \delta u^{x (0)} \Big]
-i \rho \delta F^{0x} - i \rho_d \delta B^{0x} + i \p^x \delta p 
- i \eta \Big[ \p_i \p^i \delta u^x + \frac{1}{3}\p^x \p_m \delta u^m \Big] \\ \nonumber
&-& i \zeta \p^x \p_m \delta u^m = 
i B \Big[ \rho \delta u_y - \sigma_F T \p_y\Big( \delta \big( \frac{\mu}{T} \big) \Big) + \sigma_F \Big( \delta F^{0y} - B \delta u^x \Big) \\ \nonumber
&-& \sigma_{F \tB} T \p_y\Big( \delta \big( \frac{\mu_d}{T} \big) \Big) + \sigma_{F \tB} \Big( \delta B^{0y} - \tB_{add} \delta u^x \Big)
+ \xi \p_{[z} \delta u^{x]} \Big] \\ \nonumber
&+& i \tB_{add} \Big[ \rho_d \delta u_y - \sigma_B T \p_y\Big( \delta \big( \frac{\mu_d}{T} \big) \Big) + \sigma_B \Big( \delta B^{0y} - \tB_{add} \delta u^x \Big) \\ 
&-& \sigma_{B F} T \p_y\Big( \delta \big( \frac{\mu}{T} \big) \Big) + \sigma_{B F} \Big( \delta F^{0y} - B \delta u^x \Big)
+ \xi_d \p_{[z} \delta u^{x]} \Big],\\ \nonumber
(\ep &+& p) \Big[ \omega_m  \delta u^y - i \delta u^{y (0)} \Big]
-i \rho \delta F^{0y} - i \rho_d \delta B^{0y} + i \p^y \delta p
- i \eta \Big[ \p_i \p^i \delta u^y + \frac{1}{3}\p^y \p_m \delta u^m \Big] \\ \nonumber
&-& i \zeta \p^y \p_m \delta u^m = 
i B \Big[ \rho \delta u_x - \sigma_F T \p_x \Big( \delta \big( \frac{\mu}{T} \big) \Big) + \sigma_F \Big( \delta F^{0x} + B \delta u^y \Big) \\ \nonumber
&-& \sigma_{F \tB} T \p_x \Big( \delta \big( \frac{\mu_d}{T} \big) \Big) + \sigma_{F \tB} \Big( \delta B^{0x} + \tB_{add} \delta u^y \Big)
+ \xi \p_{[y} \delta u^{z]} \Big] \\ \nonumber
&+& i \tB_{add} \Big[ \rho_d \delta u_x - \sigma_B T \p_x\Big( \delta \big( \frac{\mu_d}{T} \big) \Big) + \sigma_B \Big( \delta B^{0x} + \tB_{add} \delta u^y \Big) \\
&-& - \sigma_{B F} T \p_x\Big( \delta \big( \frac{\mu}{T} \big) \Big) + \sigma_{B F} \Big( \delta F^{0x} + B \delta u^y \Big)
+ \xi_d \p_{[y} \delta u^{z]} \Big],\\  \nonumber
(\ep &+& P) \Big[ \omega_m  \delta u^z - i \delta u^{z (0)} \Big]
-i \rho \delta F^{0z} - i \rho_d \delta B^{0z} + i \p^z \delta p
- i \eta \Big[ \p_i \p^i \delta u^z + \frac{1}{3}\p^y \p_m \delta u^m \Big] \\ 
&-& i \zeta \p^z \p_m \delta u^m = 0,\\ \nonumber
\omega_c \delta \rho &-& i \delta \rho^{(0)}  + \Big[ \omega_c \delta u_z - i \delta u_z^{(0)} \Big] \Big( \xi_B ~B + \xi_{F \tB} ~\tB_{add} \Big)
+ i \p_i \Big[
\rho \delta u^i - \sigma_F T \p^i \Big( \delta \big( \frac{\mu}{T} \big) \Big) \\ \nonumber
&+& \sigma_F  \delta F^{0i} - \sigma_{F \tB} T \p^i \Big( \delta \big( \frac{\mu_d}{T} \big) \Big) + \sigma_{F \tB}  \delta B^{0i} \Big]
+  i \p_a \Big( \frac{\xi}{2} \ep^{a jk} \p_j \delta u_k \Big)
\\ \nonumber
&+& i \p_x \Big( \sigma_F~B~\delta u^y + \sigma_{F \tB} ~\tB_{add}~\delta u^y \Big) - i \p_y \Big( \sigma_F ~B~\delta u^x 
+ \sigma_{F \tB}~\tB_{add}~\delta u^x \Big) \\ 
&+& i \p_z \Big( B ~\delta \xi_B + \tB_{add}~\delta \xi_{F \tB} \Big)
- i C_1~B~\delta E^{(F)}_ z - i C_2~\tB_{add}~\delta  \tE^{(B)}_z = 0, \\ \nonumber
\omega_{c_d} \delta \rho_d &-& i \delta \rho_d^{(0)}  + \Big[ \omega_{c_d} \delta u_z - i \delta u_z^{(0)} \Big] \Big( \xi_{\tB} ~\tB_{add} + \xi_{\tB F} ~B \Big)
 + i \p_i \Big[
\rho_d \delta u^i - \sigma_B T \p^i \Big( \delta \big( \frac{\mu_d}{T} \big) \Big) \\ \nonumber
&+& \sigma_B  \delta B^{0i} - \sigma_{B F} T \p^i \Big( \delta \big( \frac{\mu}{T} \big) \Big) + \sigma_{B F}  \delta F^{0i}
\Big]  + i \p_a \Big( \frac{\xi}{2} \ep^{a jk} \p_j \delta u_k \Big) \\ \nonumber
&+&  i \p_x \Big( \sigma_B~\tB_{add}~\delta u^y + \sigma_{BF} ~B~\delta u^y \Big) - i \p_y \Big( \sigma_B ~\tB_{add}~\delta u^x 
+ \sigma_{BF }~B ~\delta u^x \Big) \\ 
&+& i \p_z \Big( \tB_{add}~\delta \xi_\tB + B~\delta \xi_{\tB F}  \Big) 
- i C_3~B~\delta \tE^{(B) z} - i C_2~\tB_{add}~\delta E^{(F) z} = 0, 
\een
where the symbols with superscript $^{(0)}$ refer to initial values of the
perturbations. In the above relations we have denoted
\be
\omega_e = \omega + \frac{i}{\tau_e}, \qquad
\omega_m = \omega + \frac{i}{\tau_m}, \qquad
\omega_c = \omega + \frac{i}{\tau_c}, \qquad
\omega_{c_d} = \omega + \frac{i}{\tau_{c_d}}.
\ee
The Fourier transformation in the spatial directions, with the auxiliary condition that $k \rightarrow 0$, leads the following system of equations:
\ben \label{a1}
\omega_e \delta \ep &-& i \delta \ep^{(0)} - i \delta E^{(F) z} \Big(  \xi_B B + \xi_{F \tB} \tB_{add} \Big) - i \delta \tE^{(B) z} \Big( \xi_\tB  \tB_{add} 
+ \xi_{\tB F} B \Big) = 0,\\ \nonumber
(\ep &+& p) \Big[ \omega_m  \delta u^x - i \delta u^{x (0)} \Big]
-i \rho \delta F^{0x} - i \rho_d \delta B^{0x} =
 i B \Big[ \rho \delta u_y + \sigma_F \Big( \delta F^{0y} - B \delta u^x \Big) \\ \nonumber
 &+& \sigma_{F \tB} \Big( \delta B^{0y} - \tB_{add} \delta u^x \Big)
\Big] 
+ i \tB_{add} \Big[ \rho_d \delta u_y + \sigma_B \Big( \delta B^{0y} - \tB_{add} \delta u^x \Big) \\ \nonumber
&+& \sigma_{B F} \Big( \delta F^{0y} - B \delta u^x \Big) \Big],
\\ \nonumber
(\ep &+& p) \Big[ \omega_m  \delta u^y - i \delta u^{y (0)} \Big]
-i \rho \delta F^{0y} - i \rho_d \delta B^{0y} =
 i B \Big[ \rho \delta u_x+ \sigma_F \Big( \delta F^{0x} - B \delta u^y \Big) \\ \nonumber
 &+& \sigma_{F \tB} \Big( \delta B^{0x} - \tB_{add} \delta u^y \Big)
\Big] 
+ i \tB_{add} \Big[ \rho_d \delta u_x + \sigma_B \Big( \delta B^{0x} - \tB_{add} \delta u^y \Big) \\ 
&+& \sigma_{B F} \Big( \delta F^{0x} - B \delta u^y \Big) \Big],
\\ 
(\ep &+& P) \Big[ \omega_m  \delta u^z - i \delta u^{z (0)} \Big]
-i \rho \delta E^{(F) z} - i \rho_d \delta \tE^{(B) z}  = 0,\\
\omega_c \delta \rho &-& i \delta \rho^{(0)}  + \Big[ \omega_c \delta u_z - i \delta u_z^{(0)} \Big] \Big( \xi_B ~B + \xi_{F \tB} ~\tB_{add} \Big)\\ \nonumber
&-& i C_1~B~\delta E^{(F)}_ z - i C_2~\tB_{add}~\delta  \tE^{(B)}_z = 0,
\\ \label{a4}
\omega_{c_d} \delta \rho_d &-& i \delta \rho_d^{(0)}  + \Big[ \omega_{c_d} \delta u_z - i \delta u_z^{(0)} \Big] \Big( \xi_{\tB} ~\tB_{add} + \xi_{\tB F} ~B \Big)
\\ \nonumber
&-& i C_3~B~\delta \tE^{(B) z} - i C_2~\tB_{add}~\delta E^{(F) z} = 0, 
\een
In order to solve the above equations, we
write the dependence of $\delta \ep,~\delta \rho,~\delta \rho_d,~\delta p$ on $\delta \mu,~\delta \mu_d,~\delta T$ as
\ben \label{eq:257}
\delta \ep &=& e_1~\delta \mu + e_2~\delta T + e_3~\delta \mu_d 
= \Big( \frac{\p \ep}{\p \mu }\Big)_{T,\mu_d} \delta \mu +  \Big( \frac{\p \ep}{\p T} \Big)_{\mu, \mu_d} \delta T + \Big( \frac{\p \ep}{\p \mu_d} \Big)_{T,\mu} \delta \mu_d, \\
\delta \rho &=& f_1~\delta \mu + f_2~\delta T  + f_3~\delta \mu_d 
= \Big( \frac{\p \rho}{\p \mu} \Big)_{T,\mu_d} \delta \mu + \Big( \frac{\p \rho}{\p T} \Big)_{\mu,\mu_d} \delta T
+ \Big( \frac{\p \rho}{\p \mu_d} \Big)_{T,\mu} \delta \mu_d,\\
\delta \rho_d &=& g_1~\delta \mu_d + g_2~\delta T + g_3~\delta \mu = \Big( \frac{\p \rho_d}{\p \mu_d} \Big)_{T,\mu} \delta \mu_d + \Big( \frac{\p \rho_d}{\p T} \Big)_{\mu,\mu_d} \delta T
+  \Big( \frac{\p \rho_d}{\p \mu_d} \Big)_{T,\mu_d} \delta \mu,\\
\delta p &=& \rho~\delta \mu + \rho_d~\delta \mu_d + s~\delta T,
\label{eq:260}
\een
where the coefficients in the above relations depend on the details corresponding to the system in question.

We solve equations (\ref{a1})-(\ref{a4}) in terms of $\delta E^{(F)z}$ and $\delta \tE^{(B)z}$. Namely, they can be written as follows:
\ben 
\delta \mu &=&  \frac{1}{W}~\delta E^{(F)z} \Bigg[
 \frac{i}{\omega_e} \Big( \xi_B ~B + \xi_{F \tB} ~\tB_{add} \Big)\Big(f_2 g_1 - g_2 f_3 \Big) \\ \nonumber
 &+& 
\bigg( - \frac{\rho}{\ep+p} \frac{i}{\omega_m} \Big(\xi_B~B + \xi_{F \tB}~\tB_{add} \Big)
 + \frac{i}{\omega_c} C_1 B \bigg)  \Big( g_2 e_3 - e_2 g_1 \Big) \\ \nonumber
&+&\bigg( - \frac{\rho}{\ep+p} \frac{i}{\omega_m}  \Big(\xi_\tB~\tB_{add} + \xi_{\tB F}~B \Big)
+ \frac{i}{\omega_{cd}} C_4 \tB_{add} \bigg) \Big( f_3 e_2 - f_2 e_3 \Big)
\Bigg] \\ \nonumber
&+&  \frac{1}{W}~\delta \tE^{(B)z} \Bigg[
\frac{i}{\omega_e} \Big( \xi_\tB ~\tB_{add} + \xi_{ \tB F} ~B \Big) \Big(f_2 g_1 - g_2 f_3 \Big) \\ \nonumber
 &+& 
\bigg( - \frac{\rho_d}{\ep+p} \frac{i}{\omega_m} \Big(\xi_B~B + \xi_{F \tB}~\tB_{add} \Big)
 + \frac{i}{\omega_c } C_2 \tB_{add} \bigg) \Big( g_2 e_3 - e_2 g_1 \Big) \\ \nonumber
&+&\bigg( - \frac{\rho_d}{\ep+p} \frac{i}{\omega_m}  \Big(\xi_\tB~\tB_{add} + \xi_{\tB F}~B \Big)
+ \frac{i}{\omega_{cd}} C_3 B \bigg) \Big( f_3 e_2 - f_2 e_3 \Big)
\Bigg] + \dots \\ \nonumber
&=& M_1\delta E^{(F)z} +M_2\delta \tE^{(B)z} + \dots,
\een

\ben 
\delta \mu_d &=&  \frac{1}{W}~\delta E^{(F)z} \Bigg[
\frac{i}{\omega_{e}} \Big( \xi_B ~B + \xi_{F \tB} ~\tB_{add} \Big)  \Big(f_1 g_2 - f_2 g_3 \Big) \\ \nonumber
 &+& 
\bigg( - \frac{ \rho}{\ep+p} \frac{i}{\omega_m} \Big(\xi_B~B + \xi_{F \tB}~\tB_{add} \Big)
 + \frac{i}{\omega_c} C_1 B \bigg) \Big( e_2 g_3 - g_2 e_1 \Big) \\ \nonumber
&+&\bigg( - \frac{\rho}{\ep+p} \frac{i}{\omega_m}  \Big(\xi_\tB~\tB_{add} + \xi_{\tB F}~B \Big)
+ \frac{i}{\omega_{cd}} C_4 \tB_{add} \bigg) \Big( e_1 f_2 - e_2 f_1 \Big)
\Bigg] \\ \nonumber
&+&  \frac{1}{W}~\delta \tE^{(B)z} \Bigg[
\frac{i}{\omega_{e}} \Big( \xi_\tB ~\tB_{add} + \xi_{ \tB F} ~B \Big) \Big(f_1 g_2 - f_2 g_3 \Big) \\ \nonumber
 &+& 
\bigg( - \frac{ \rho_d}{\ep+p} \frac{i}{\omega_m} \Big(\xi_B~B + \xi_{F \tB}~\tB_{add} \Big)
 + \frac{i}{\omega_c} C_2 \tB_{add} \bigg) \Big( g_3 e_2 - e_1 g_2 \Big) \\ \nonumber
&+&\bigg( - \frac{\rho_d}{\ep+p} \frac{i}{\omega_m}  \Big(\xi_\tB~\tB_{add} + \xi_{\tB F}~B \Big)
+ \frac{i}{\omega_{cd}} C_3 B \bigg) \Big( e_1 f_2 - e_2 f_1 \Big)
\Bigg] + \dots \\ \nonumber
&=& D_1\delta E^{(F)z} +D_2\delta \tE^{(B)z} + \dots,
\een

\ben
\delta T &=&  \frac{1}{W}~\delta E^{(F)z} \Bigg[
\frac{i}{\omega_e} \Big( \xi_B ~B + \xi_{F \tB} ~\tB_{add} \Big) \Big(f_3 g_3 - f_1 g_1 \Big) \\ \nonumber
 &+& 
\bigg( - \frac{\rho}{\ep+p} \frac{i}{\omega_m} \Big(\xi_B~B + \xi_{F \tB}~\tB_{add} \Big)
 + \frac{i}{\omega_c} C_1 B \bigg) \Big( e_1 g_1 - g_3 e_3 \Big) \\ \nonumber
&+&\bigg( - \frac{\rho}{\ep+p} \frac{i}{\omega_m}  \Big(\xi_\tB~\tB_{add} + \xi_{\tB F}~B \Big)
+ \frac{i}{\omega_{cd}} C_4 \tB_{add} \bigg) \Big( e_3 f_1 - e_1 f_3 \Big)
\Bigg] \\ \nonumber
&+&  \frac{1}{W}~\delta \tE^{(B)z} \Bigg[
\frac{i}{\omega_e} \Big( \xi_\tB ~\tB_{add} + \xi_{ \tB F} ~B \Big) \Big(f_3 g_3 - f_1 g_1 \Big) \\ \nonumber
 &+& 
\bigg( - \frac{ \rho_d}{\ep+p} \frac{i}{\omega_m} \Big(\xi_B~B + \xi_{F \tB}~\tB_{add} \Big)
 + \frac{i}{\omega_c} C_2 \tB_{add} \bigg) \Big( g_1 e_1 - e_3 g_3 \Big) \\ \nonumber
&+&\bigg( - \frac{\rho_d}{\ep+p} \frac{i}{\omega_m}  \Big(\xi_\tB~\tB_{add} + \xi_{\tB F}~B \Big)
+ \frac{i}{\omega_{cd}} C_3 B \bigg) \Big( e_3 f_1 - e_1 f_3 \Big)
\Bigg] + \dots \\ \nonumber
&=& T_1\delta E^{(F)z} +T_2\delta \tE^{(B)z} + \dots,
\een
\be
\delta u^z = \frac{\rho}{\ep + p} \frac{i}{\omega_m} \delta E^z + \frac{\rho_d}{\ep + p} \frac{i}{\omega_m} \delta \tE^z + \dots,
\ee
where we have denoted by $W$
\be
W = det \left(\begin{array}{ccc}  e_1 &  e_2 &  e_3 \\
 f_1 &  f_2 &  f_3 \\
 g_3 &  g_2 &  g_1 \end{array} \right).
\ee


By $'\dots'$ one has in mind the terms which are unrelated to $ \delta E^{(F)z}$ and $ \delta \tE^{(B)z}$, which disappear when 
we choose the initial values of the other perturbations equal to zero.
In what follows, for the brevity of notation, one assigns all terms standing in front of $ \delta E^z$ and $ \delta \tE^z$, respectively by
\ben
\delta \mu &=& M_1~\delta E^{(F)z} + M_2~\delta \tE^{(B)z},\\
\delta \mu_d &=& D_1~\delta E^{(F)z} + D_2~\delta \tE^{(B)z},\\
\delta T &=& T_1~\delta E^{(F)z} + T_2~\delta \tE^{(B)z}.
\een
They will be needed for substituting them into the perturbation relations connected with $z$-directed gauge field currents. On this account we have
\ben \nonumber
\delta j^z(F) &=& \rho~ \delta u^z - \sigma_F ~T ~\p^z\Big( \delta \frac{\mu}{T} \Big) + \sigma_F ~\delta E^{(F)z} 
- \sigma_{F \tB} ~T ~\p^z\Big( \delta \frac{\mu_d}{T} \Big) + \sigma_{F \tB} ~\delta \tE^{(B)z} \\ 
&+& \delta \xi_B~ B +  \delta \xi_{F \tB} ~\tB_{add} + \xi ~\p_{[x} \delta u_{y]},
\een
and for $B_{\mu \nu}$ field current it yields
\ben \nonumber
\delta j^z(B) &=& \rho_d~ \delta u^z - \sigma_B~T ~\p^z\Big( \delta \frac{\mu_d}{T} \Big) + \sigma_B ~\delta \tE^{(B)z} 
- \sigma_{B F} ~T ~\p^z\Big( \delta \frac{\mu}{T} \Big) + \sigma_{BF} ~\delta E^{(F)z} \\ 
&+& \delta \xi_\tB~ \tB_{add} +  \delta \xi_{\tB F} ~B + \xi ~\p_{[x} \delta u_{y]}.
\een
It can be verified that in the limit of $k \rightarrow 0$ they reduce to
\be
\delta j^z(F) = \rho~ \delta u^z + \sigma_F ~\delta E^{(F)z} 
+ \sigma_{F \tB} ~\delta \tE^{(B)z} \\ 
+ \delta \xi_B~ B +  \delta \xi_{F \tB} ~\tB_{add},
\ee
for the perturbations of the Maxwell field current, and for the additional one we obtain
\be
\delta j^z(B) = \rho_d~ \delta u^z + \sigma_B ~\delta \tE^{(B)z}  + \sigma_{BF} ~\delta E^{(F)z} 
+ \delta \xi_\tB~ \tB_{add} +  \delta \xi_{\tB F} ~B.
\ee
Inserting the explicit values for $\delta \xi_B,~\delta \xi_{F \tB}$ and $\delta \xi_\tB,~ \delta \xi_{\tB F} $, 
obtained  from (\ref{eq269})-(\ref{eq270}) with help of (\ref{eq:257})-(\ref{eq:260}), 
one arrives at the relation
\ben 
\delta j_z(F) &=&  \delta E^{(F)}_z \Bigg[
\sigma_F + \frac{\rho^2}{\ep+p}~ \frac{i}{\omega_m} + B ~\Big( M_1~H_1 + D_1~H_2 + T_1~H_3\Big)  \\ \nonumber
&+& 
\tB_{add}~ \Big( D_1~G_1 + M_1~G_2 + T_1~G_3\Big) \Bigg] \\ \nonumber
&+& \delta \tE^{(B)}_z \Bigg[ \sigma_{F \tB} + 
\frac{\rho \rho_d}{\ep + p} ~\frac{i}{\omega_m} + B \Big( M_2~H_1 + D_2~H_2 + T_2~H_3\Big)  \\ \nonumber
&+& \tB_{add}~ \Big( D_2~G_1 + M_2~G_2 + T_2~G_3\Big) 
\Bigg],
\een
where we have defined the following quantities connected with the terms multiplied by $B$-magnetic field:
\ben
H_1 &=& C_1 \Big( 1 - \frac{\rho \mu}{\ep + p} \Big) + \frac{x_1}{2(\ep + p)} \Big( -f_1 + \frac{\rho}{\ep + p} (e_1 + \rho) \Big),\\
H_2 &=& - C_3 ~\frac{\rho \mu_d}{\ep + p} + \frac{x_1}{2(\ep + p)} \Big( -f_3 + \frac{\rho}{\ep + p} (e_3 + \rho_d) \Big),\\
H_3 &=& - \frac{2 \rho T}{\ep + p} \tga_1 + \frac{x_1}{2(\ep + p)} \Big( -f_2+ \frac{\rho}{\ep + p} (e_2+ s) \Big),\\
x_1 &=& C_3~\mu_d^2 + C_1~\mu^2 + 2~\tga_1~T^2,
\een
and with the terms multiplied by $\tB_{add}$
\ben
G_1 &=& C_2 \Big( 1 - \frac{\rho \mu}{\ep + p} \Big) + \frac{x_2}{(\ep + p)^2}\Big( \rho (\rho_d  + e_3 ) - f_3 (\ep +p) \Big),\\
G_2 &=& - C_2 \frac{\rho \mu_d}{\ep + p}  + \frac{x_2}{(\ep + p)^2}\Big( \rho( \rho + e_1 ) - f_1 (\ep +p) \Big),\\
G_3 &=& - \frac{2 \rho T}{\ep + p} \tga_2 + \frac{x_2}{(\ep + p)^2}\Big( \rho (s + e_2) - f_3 (\ep +p) \Big),\\
x_2 &=& C_2~\mu~\mu_d +\tga_2~T^2.
\een

On the other hand, for the second gauge field current one obtains
\ben 
\delta j_z(B) &=&  \delta E^{(F)}_z \Bigg[
\sigma_{BF} + \frac{\rho \rho_d}{\ep +p}~ \frac{i}{\omega_m} + \tB_{add}~ \Big( M_1~I_1 + D_1~I_2 + T_1~I_3\Big) \\ \nonumber
&+& B ~\Big( M_1~J_1 + D_1~J_2 + T_1~J_3\Big) \Big] \\ \nonumber
&+& \delta \tE^{(B)}_z \Bigg[
\sigma_B + \frac{\rho_d^2}{\ep +p} \frac{i}{\omega_m} + \tB_{add}~ \Big( M_2~I_1 + D_2~I_2 + T_2~I_3\Big) \\ \nonumber
&+& B~ \Big( M_2~J_1 + D_2~J_2 + T_2~J_3\Big) 
\Bigg],
\een
where we have defined the following quantities:
\ben
I_1 &=& C_2 \Big( 1 - \frac{\rho_d \mu_d}{\ep + p} \Big) + \frac{x_2}{(\ep + p)^2} \Big( \rho_d (\rho + e_1 ) - g_3 (\ep +p) \Big),\\
I_2 &=& - C_2 ~\frac{\rho \mu_d}{\ep + p} + \frac{x_2}{(\ep + p)^2} \Big( \rho_d (\rho_d + e_3 ) - g_1(\ep +p) \Big),\\
I_3 &=& - \frac{2 \rho_d T}{\ep + p} \tga_2 + \frac{x_2}{(\ep + p)^2} \Big( \rho_d (s + e_2 ) - g_2(\ep +p) \Big),
\een
and
\ben
J_1 &=& - C_1 \frac{\rho_d  \mu}{\ep + p}  +  \frac{x_1}{2(\ep + p)} \Big( -g_3 + \frac{\rho_d}{\ep + p} (e_1 + \rho) \Big),\\
J_2 &=&  C_3 \Big( 1 - \frac{\rho_d \mu}{\ep + p}  \Big) +  \frac{x_1}{2(\ep + p)} \Big( -g_1 + \frac{\rho_d}{\ep + p} (e_3 + \rho_d) \Big),\\
J_3 &=& - \frac{2 \rho_d T}{\ep + p} \tga_1+ \frac{x_1}{2(\ep + p)} \Big( -g_2 + \frac{\rho_d}{\ep + p} (e_2 + s) \Big).
\een
Consequently,  we can write them in a more compact forms, given by
\ben \label{cur1}
\delta j_z(F) &=&  {\tilde \sigma}_F~\delta E^{(F)}_z + {\tilde \sigma}_{F B}~ \delta \tE^{(B)}_z,\\ \label{cur2}
\delta j_z(B) &=&  {\tilde \sigma}_B~\delta \tE^{(B)}_z + {\tilde \sigma}_{ B F}~ \delta E^{(F)}_z.
\een
By virtue of the equations (\ref{cur1})-(\ref{cur2}), we can read off the explicit forms of the adequate conductivities 
\ben \label{sf}
 {\tilde \sigma}_F &=& 
 \sigma_F + \frac{\rho^2}{\ep+p}~ \frac{i}{\omega_m} + B ~\Big( M_1~H_1 + D_1~H_2 + T_1~H_3\Big) \\ \nonumber
&+& \tB_{add}~ \Big( D_1~G_1 + M_1~G_2 + T_1~G_3\Big),\\
{\tilde \sigma}_{F B} &=& 
\sigma_{F \tB} + 
\frac{\rho \rho_d}{\ep + p} ~\frac{i}{\omega_m} + B \Big( M_2~H_1 + D_2~H_2 + T_2~H_3\Big)  \\ \nonumber
&+& \tB_{add}~ \Big( D_2~G_1 + M_2~G_2 + T_2~G_3\Big),\\ 
 {\tilde \sigma}_B &=& 
 \sigma_{B} + \frac{\rho_d^2}{\ep +p} \frac{i}{\omega_m} + \tB_{add}~ \Big( M_2~I_1 + D_2~I_2 + T_2~I_3\Big) \\ \nonumber
&+& B~ \Big( M_2~J_1 + D_2~J_2 + T_2~J_3\Big),\\  \label{sbf}
 {\tilde \sigma}_{ B F} &=& 
 \sigma_{BF} + \frac{\rho \rho_d}{\ep +p}~ \frac{i}{\omega_m} + \tB_{add}~ \Big( M_1~I_1 + D_1~I_2 + T_1~I_3\Big) \\ \nonumber
&+& B ~\Big( M_1~J_1 + D_1~J_2 + T_1~J_3\Big).
  \een
These results constitute the expressions for the longitudinal electric conductivities of a fluid with chiral anomaly and $\zz$ topological charge, in the background
of the adequate magnetic fields $B$ and $\tB_{add}$, connected with Maxwell and auxiliary $U(1)$-gauge fields.

\subsection{The limits}
In this subsection we shall discuss physical consequences of the model. It is important to note that
DC conductivity corresponding to $\omega=0$ limit is finite, if all of the relaxation 
times $\tau_i$, i.e., energy for $i=e$, momentum ($i=m$) and and two charge relaxations ($i=c,cd$), are finite. 
This agrees with the earlier observation \cite{lan15} for the model without $\zz$ topological charge. 
For the finite values of the relaxation times, the conductivities 
have both real $\sigma'(\omega)$ and imaginary $\sigma''(\omega)$ parts. The imaginary one characterizes by an antisymmetric contribution
in frequency, resulting from the terms like $i\omega/(\omega^2+1/\tau_i^2)$. The real parts
of the conductivities are governed by the corresponding relaxation times and contain terms with
factors $(1/\tau_i)/(\omega^2+1/\tau_i^2)$, which in the limit of large relaxation times (i.e., $1/\tau \rightarrow 0$)
envisage the zero frequency, $\delta(\omega)$, contribution only.

If we consider the case without relaxation terms, i.e., $\omega_e,~\omega_m,~\omega_c,~\omega_{cd} \rightarrow \omega$ 
should be replaced in
 (\ref{sf})-(\ref{sbf}). The conductivities will have poles in the imaginary part at $\omega=0$. 
Consequently, one will encounter $\delta (\omega)$ in the real part of the DC-conductivities in question.
For a system without $\zz$ topological charge there exist only a single non-zero
parameter $C_1=C$ and our general formula for the conductivity matrix reduces to the element $\tilde\sigma_F$, which agrees with that found earlier \cite{lan15}.

Because of the complexity of the obtained results we pay attention to some simpler limit cases. Namely, 
our analysis will be addressed to the case when
$\rho = \mu = 0$, which implies also that $f_2 = f_3 = 0$. In the case under consideration, one obtains the following forms of the conductivities:
\ben
 {\tilde \sigma}_F \Big(\rho=\mu=0 \Big) &=& \sigma_F + B \Big[
 \frac{i}{\omega_c} \Big(
 \frac{C_1^2~B}{f_1} - \frac{C_1~B}{2 (\ep + p)} \big(C_3~\mu_d^2 + 2 \tga_1 T^2 \big) \Big) \Big] \\ \nonumber
 &-&
 \tB_{add}~ \frac{i}{\omega_c} ~\frac{C_1~B~\tga_2~T^2}{\ep +p},\\
 {\tilde \sigma}_{F B} \Big(\rho=\mu=0 \Big) &=& \sigma_{F \tB} + B~M_2^{(L)} H_1^{(L)} + \tB_{add} \Big( D_2^{(L)} G_1^{(L)} + M_2^{(L)} C_2 \Big),\\
  {\tilde \sigma}_{BF} \Big(\rho=\mu=0 \Big) &=& \sigma_{BF} +
  \tB_{add}~\Big[ \frac{i}{\omega_e} ~\frac{C_1~B}{f_1} ~I_1^{(L)} + D_1^{(L)} I_2^{(L)}  + T_1^{(L)} I_3^{(L)} \Big] \\ \nonumber
 & +& B~\Big[ \frac{i}{\omega_e} ~\frac{C_1~B}{f_1} ~J_1^{(L)} + D_1^{(L)} J_2^{(L)}+ T_1^{(L)} I_2^{(L)} + T_1^{(L)} I_3^{(L)} \Big],\\
  {\tilde \sigma}_{ B } \Big(\rho=\mu=0 \Big) &=& \sigma_B
  + \frac{\rho_d^2}{\ep + p} \frac{i}{\omega_e} + \tB_{add} \Big[
  M_2^{(L)} I_1^{(L)} + D_2^{(L)} I_2^{(L)} + T_2^{(L)} I_3^{(L)} \Big] \\ \nonumber
  &+& B~\Big[ M_2^{(L)} J_1^{(L)} + D_2^{(L)} J_2^{(L)} + T_2^{(L)} J_3^{(L)} \Big],
 \een
where for the brevity of the subsequent notion we have denoted the following quantities:
\be
M_2^{(L)} = - \frac{\rho_d}{\ep + p} \frac{i}{\omega_e} ~\frac{\tB_{add} C_2~\mu_d}{f_1} + \frac{i}{\omega_c} ~\frac{\tB_{add} C_2~\mu_d}{f_1} ,
\ee
\be
D_1^{(L)} = \frac{i}{\omega_e} \frac{\tB_{add} C_2 \mu_d ~g_2}{g_2 e_3 - e_2 g_1} + \frac{i}{\omega_c} \frac{C_1 B ~(g_3 e_2-g_2 e_1)}{f_1 (g_2 e_3 - e_2 g_1)}
- \frac{i}{\omega_{cd}} \frac{C_4 \tB_{add} e_2}{g_2 e_3-e_2 g_1},
\ee
\ben
D_2^{(L)} &=&\frac{i}{\omega_e} \frac{\tB_{add} C_2 \mu_d ~g_2}{g_2 e_3 - e_2 g_1} - \frac{i}{\omega_m} \frac{\rho_d}{\ep +p}
\frac{C_2 \tB_{add} \mu_d ~(g_3 e_2-g_2 e_1)}{f_1 (g_2 e_3 - e_2 g_1)}
- \frac{i}{\omega_{cd}} \frac{C_3 \tB_{add} e_2}{g_2 e_3-e_2 g_1}\\ \nonumber
&+& \frac{i}{\omega_m} \frac{\rho_d}{\ep +p}
\frac{C_2 \tB_{add} \mu_d ~e_2}{f_1 (g_2 e_3 - e_2 g_1)},
\een
\be
T_1^{(L)} = - \frac{i}{\omega_e} \frac{\tB_{add} C_2 \mu_d~g_1}{g_2 e_3 - e_2 g_1} + \frac{i}{\omega_c} \frac{C_1 B (e_1 g_1- e_3 g_3)}{f_1  (g_2 e_3 - e_2 g_1)},
\ee
\ben
T_2^{(L)} &=& - \frac{i}{\omega_e} \frac{\tB_{add} C_2 \mu_d~g_1}{g_2 e_3 - e_2 g_1} - 
\frac{i}{\omega_m}  \frac{\rho_d}{\ep + p}\frac{\tB_{add} C_2 \mu_d~(e_1g_1- e_3 g_3)}{f_1( g_2 e_3 - e_2 g_1)}\\ \nonumber
& -& 
\frac{i}{\omega_{cd}} \frac{C_3 \tB_{add} e_3}{g_2 e_3-e_2 g_1}
- \frac{i}{\omega_m}  \frac{\rho_d}{\ep + p}\frac{\tB_{add} C_2 \mu_d~ e_3}{ g_2 e_3 - e_2 g_1},\\ 
H_1^{(L)} &=& C_1 - \frac{C_3 \mu_d^2 + 2 \tga_1 T^2}{2(\ep + p)} f_1,\\
G_2^{(L)} &=& -f_1 \frac{\tga_2 T^2}{\ep + p},\\
I_1^{(L)} &=& C_1 \Big(1 - \frac{\rho_d \mu_d}{\ep + p} \Big) + \frac{\tga_2 T^2}{(\ep + p)^2} \Big( e_1 \rho_d - g_3 (\ep + p) \Big),\\
I_2^{(L)} &=& \frac{\tga_2 T^2}{(\ep + p)^2}\Big( \rho_d (\rho_d + e_3 ) - g_1(\ep +p) \Big),\\
I_3^{(L)} &=& - 2 \frac{\rho_d \tga_2 T}{\ep + p} + \frac{\tga_2 T^2}{(\ep + p)^2} \Big( \rho_d (s + e_2 ) - g_2(\ep +p) \Big),\\
J_1^{(L)} &=& \frac{C_3 \mu_d^2 + 2 \tga_1 T^2}{2(\ep + p)} \Big( -g_3 + \frac{\rho_d e_1}{\ep + p}  \Big),\\
J_2^{(L)} &=&C_3 + \frac{C_3 \mu_d^2 + 2 \tga_1 T^2}{2(\ep + p)} \Big( -g_1 + \frac{\rho_d ( e_3 + \rho_d)}{\ep + p}  \Big),\\
J_3^{(L)}  &=& - 2 \frac{\rho_d \tga_1 T}{\ep + p} + \frac{C_3 \mu_d^2 + 2 \tga_1 T^2}{2(\ep + p)} \Big( -g_2 + \frac{\rho_d (e_2 + s)}{\ep + p}  \Big).
\een
The above relations simplify to great extend, when one assumes that the magnetic fields are equal 
to zero. Consequently this assumption leads to
\ben
{\tilde \sigma}_{F} \Big(\rho=\mu=0 \Big) &=& \sigma_F, \\
{\tilde \sigma}_{FB} \Big(\rho=\mu=0 \Big) &=& \sigma_{F \tB},\\
 {\tilde \sigma}_{BF} \Big(\rho=\mu=0 \Big) &=& \sigma_{BF},\\
{\tilde \sigma}_{B} \Big(\rho=\mu=0 \Big) &=& \sigma_{B} + \frac{i}{\omega_m}\frac{\rho_d^2}{\ep + p} .
\een
One can see that in the considered case the additional $U(1)$-gauge field density $\rho_d$ gives the finite conductivity.

On the other hand, in the limit when $\rho_d = \mu_d = 0$, which implies that $g_2 = g_3 = 0$, the relations for the conductivities yield
\ben \nonumber
{\tilde \sigma}_F \Big(\rho_d=\mu_d=0 \Big) &=& \sigma_F + \frac{\rho^2}{\ep + p} \frac{i }{\omega_m}
+ B~ \Big[ M_1^{(D)} H_1^{(D)} + D_1^{(D)} H_2^{(D)} + T_1^{(D)} H_3^{(D)} \Big] \\
&=& \tB_{add}~\Big[ D_1^{(D)} G_1^{(D)} + M_1^{(D)} G_2^{(D)} + T_1^{(D)} G_3^{(D)} \Big],\\
{\tilde \sigma}_{F B} \Big(\rho_d=\mu_d=0 \Big) &=& \sigma_{F \tB} + + B~ \Big[ M_2^{(D)} H_1^{(D)} + \frac{i}{\omega_{cd}} \frac{C_3 B}{g_1}H_2^{(D)} 
+ T_2^{(D)} H_3^{(D)} \Big] \\
&+& \tB_{add}~\Big[ \frac{i}{\omega_{cd}} \frac{C_3 B}{g_1}
G_1^{(D)} + M_2^{(D)} G_2^{(D)} + T_2^{(D)} G_3^{(D)} \Big],\\
 {\tilde \sigma}_{BF} \Big(\rho_d=\mu_d=0 \Big) &=& \sigma_{BF} + \tB_{add} \Big[
 M_1^{(D)} C_2 - D_1^{(D)} \frac{\tga_2 T^2 g_1}{\ep + p} \Big] \\
 &+& B~D_1^{(D)} \Big( C_3 - \frac{C_1 \mu^2 + 2 \tga_1 T^2}{2(\ep + p)} g_1 \Big), \\
 {\tilde \sigma}_{ B} \Big(\rho_d=\mu_d=0 \Big) &=& \sigma_B + \tB_{add} 
 \Big[
 M_2^{(D)} C_2 - \frac{i}{\omega_m} \frac{C_3 B}{g_1} \frac{\tga_2 T^2 g_1}{\ep + p} \Big] \\ \nonumber
 &+& B~\frac{i}{\omega_{cd}} \frac{C_3 B}{g_1} \Big(
 C_3 - \frac{C_1 \mu^2 + 2 \tga_1 T^2}{2(\ep + p)} g_1 \Big).
\een
In the above relations we set the quantities
\ben
M_1^{(D)} &=& \Big(\frac{i}{\omega_e} f_2 + \frac{\rho}{\ep + p}\frac{i}{\omega_m}\Big) \frac{1}{e_1f_2-e_2f_1}\Big(1 - \frac{1}{g_1}\Big) \Big[
C_1 \mu B \Big(1 - \frac{1}{2} \frac{\rho \mu}{\ep + p} \Big) \\ \nonumber
&-& \frac{\rho \tga_1 T^2 B}{e\ + p}  - \frac{\rho \tga_2 T^2 \tB_{add}}{\ep + p}\Big] 
- \frac{i}{\omega_c}\frac{C_1 B e_2}{e_1f_2-e_2f_1} + \frac{i}{\omega_{cd}}\frac{C_4 \tB_{add} (e_2 f_3-e_3f_2)}{g_1(e_1f_2-e_2f_1)},
\een
\be
M_2^{(D)} = \frac{1}{e_1f_2-e_2f_1} \Big[\frac{i}{\omega_e}  \tB_{add} C_2 \mu f_2 - \frac{i}{\omega_c} C_2 \tB_{add} e_2 +
\frac{i}{\omega_{cd}} C_3 B (e_2f_3 - e_3f_2) \Big],
\ee
\be
D_1^{(D)} =-\frac{i}{\omega_m} \frac{\rho}{\ep + p}~\frac{\tB_{add} C_2 \mu }{g_1} + \frac{i}{\omega_{cd}} \frac{C_4 \tB_{add}}{g_1},
\ee
\ben
T_1^{(D)} &=& \Big[
C_1 \mu B \Big(1 - \frac{1}{2} \frac{\rho \mu}{\ep + p} \Big) 
- \frac{\rho \tga_1 T^2 B}{e\ + p}  - \frac{\rho \tga_2 T^2 \tB_{add}}{\ep + p}\Big] \frac{1}{e_1f_2 -e_2f_1} \\
\Big( &-&\frac{i}{\omega_e} f_1 
- \frac{i}{\omega_m} \frac{\rho}{\ep+p} \Big(e_1 + \frac{f_1e_3 -f_3e_1}{g_1} \Big) \Big)\\
&+& \frac{1}{e_1f_2 -e_2f_1} \Big[ \frac{i}{\omega_c} C_1 B e_1 + \frac{i}{\omega_{cd}} C_4 \tB_{add} \frac{f_1e_3 - f_3 e_1}{g_1} \Big],
\een 
\be
T_2^{(D)} = \frac{1}{e_1f_2 -e_2f_1} \Big[ -\frac{i}{\omega_e} \tB_{add} C_2 \mu f_1 + -\frac{i}{\omega_c} C_2 \tB_{add} e_1 +
-\frac{i}{\omega_{cd}} C_3 B \frac{f_1e_3 - f_3e_1}{g_1} \Big],
\ee
\be
H_1^{(D)} = C_1 \Big( 1 - \frac{\rho \mu}{\ep + p} \Big) + \frac{C_1 \mu + 2 \tga_1 T^2}{2(\ep +p)}\Big( -f_1 + \frac{\rho}{\ep+p} (e_1+\rho) \Big),
\ee
\be
H_2^{(D)} = \frac{C_1 \mu^2 + 2 \tga_1 T^2}{2(\ep + p)} \Big(-f_3 + \frac{\rho e_3}{\ep +p} \Big),
\ee
\be
H_3^{(D)} = - 2 \frac{\rho \tga_1 T}{\ep + p} + \frac{C_1 \mu + 2 \tga_1 T^2}{2(\ep +p)}\Big(-f_2 + \frac{\rho}{\ep+p} (e_2 + s) \Big),
\ee
\be
G_1^{(D)} = C_2 \Big( 1 -\frac{\rho \mu}{\ep+p} \Big) + \frac{\tga_2 T^2}{(\ep +p)^2} \Big( e_3 \rho - f_3 (\ep +p) \Big),
\ee
\be
G_2^{(D)} = \frac{\tga_2 T^2}{(\ep +p)^2} \Big( \rho^2 - f_1 (\ep +p) \Big),
\ee
\be
G_3^{(D)} = -2 \frac{\rho \tga_2 T^2}{\ep + p} + \frac{\tga_2 T^2}{(\ep +p)^2} \Big( s \rho - f_3 (\ep +p) \Big).
\ee
As in the latter case, let us suppose that the magnetic fields are equal to zero. It implies the following:
\ben
{\tilde \sigma}_{F} \Big(\rho_d=\mu_d=0 \Big) &=& \sigma_F + \frac{i}{\omega_m} \frac{\rho^2}{\ep +p}, \\
{\tilde \sigma}_{FB} \Big(\rho_d=\mu_d=0 \Big) &=& \sigma_{F \tB},\\
 {\tilde \sigma}_{BF} \Big(\rho_d=\mu_d=0 \Big) &=& \sigma_{BF},\\
{\tilde \sigma}_{B} \Big(\rho_d=\mu_d=0 \Big) &=& \sigma_{B}.
\een
As was previously stated, the additional gauge field density, now $\rho$, gives the finite conductivity.

In the case when $\rho = \mu =\rho_d = \mu_d = 0$, and $f_2 = f_3 =g_2 = g_3 = 0$, one receives the relations provided by
\be
{\tilde \sigma}_F \Big(\rho_m=\mu_m=0 \Big) = \sigma_F + \frac{i}{\omega_c} \frac{C_1^2 B^2}{f_1} + \frac{i}{\omega_{cd}} \frac{C_2 C_4 \tB_{add}}{g_1}
- \frac{i}{\omega_c}\frac{C_1 B}{\ep + p} T^2 \Big( \tga_1 B + \tga_2 \tB_{add} \Big),
\ee 
\ben
{\tilde \sigma}_{FB} \Big(\rho_m=\mu_m=0 \Big) = \sigma_{F \tB} &+& 
\frac{i}{\omega_c} \frac{C_1 C_2 B \tB_{add}}{f_1} + \frac{i}{\omega_{cd}} \frac{C_2 C_3 B \tB_{add}}{g_1} \\ \nonumber
&-& \frac{i}{\omega_{c}} \frac{C_2 \tB_{add}}{\ep +p}
T^2 \Big( \tga_1 B + \tga_2 \tB_{add} \Big),
\een
\ben
{\tilde \sigma}_{BF} \Big(\rho_m=\mu_m=0 \Big) = \sigma_{BF} &+& \frac{i}{\omega_c} \frac{C_1 C_2 B \tB_{add}}{f_1}
+ \frac{i}{\omega_{cd}} \frac{C_3 C_4 B \tB_{add}}{g_1}  \\ \nonumber
&-& \frac{i}{\omega_{cd}}\frac{C_4 \tB_{add}}{\ep +p}  T^2 \Big( \tga_1 B + \tga_2 \tB_{add} \Big),
\een
\ben
{\tilde \sigma}_{B} \Big(\rho_m=\mu_m=0 \Big) = \sigma_{B} &+& \frac{i}{\omega_c} \frac{C_2^2 B \tB_{add}}{f_1}
+ \frac{i}{\omega_{cd}} \frac{C_3^2 B^2 }{g_1}  \\ \nonumber
&-& \frac{i}{\omega_{cd}}\frac{C_3  B}{\ep +p}  T^2 \Big( \tga_1 B + \tga_2 \tB_{add} \Big),
\een
where for the brevity of the notion we introduce index 'm' denoting both cases, i.e., the case of the ordinary Maxwell field and the additional $U(1)$-gauge one.

Neglecting the presence of $\zz$ topological charge, which means that $\tB_{add} =0$, 
and assuming $\tga_1=0$, we get 
\ben \label{sf-add}
{\tilde \sigma}_{F} \Big(\rho_m=\mu_m=0 \Big) &=& \sigma_F + \frac{i}{\omega_c} \frac{C_1^2~B^2}{\Big( \frac{\p \rho}{\p \mu} \Big)_{ \mid_{T, \mu_d}}},\\
{\tilde \sigma}_{FB} \Big(\rho_m=\mu_m=0 \Big) &=& \sigma_{F \tB},\\
 {\tilde \sigma}_{BF} \Big(\rho_m=\mu_m=0 \Big) &=& \sigma_{BF},\\ \label{sbf-add}
{\tilde \sigma}_{B} \Big(\rho_m=\mu_m=0 \Big) &=& \sigma_{B} + \frac{i}{\omega_{cd}} \frac{C_3^2~B^2}{\Big( \frac{\p \rho_d}{\p \mu_d} \Big)_{ \mid_{T, \mu}}}.
\een
From the above relations one can draw a conclusion that even at zero densities we obtain still finite DC-conductivities. The terms in question can only be dissipated by 
the charge dissipations, $\omega_c$ and $\omega_{cd}$, respectively. 
 On the other hand, the form of ${\tilde \sigma}_F $ conductivity is the same as in \cite{lan15} (see the next section), but in our case $\rho$ is also dependent on the auxiliary gauge field components.

Finally, we assume that $B=T=0$. This limit is beyond the hydrodynamical calculations which require
$B/T^2 \ll 1$, as was mentioned in subsection 2.1, but for the completeness of the considerations we present these relations
\ben \label{sdf}
{\tilde \sigma}_{F} \Big(\rho_m=\mu_m=0 \Big) &=& \sigma_F + \frac{i}{\omega_{cd}} \frac{C_2 C_4~\tB_{add}^2}{\Big( \frac{\p \rho_d}{\p \mu_d} \Big)_{ \mid_{T, \mu}}},\\
{\tilde \sigma}_{FB} \Big(\rho_m=\mu_m=0 \Big) &=& \sigma_{F \tB},\\
 {\tilde \sigma}_{BF} \Big(\rho_m=\mu_m=0 \Big) &=& \sigma_{BF},\\ \label{sdbf}
{\tilde \sigma}_{B} \Big(\rho_m=\mu_m=0 \Big) &=& \sigma_{B} + \frac{i}{\omega_{c}} \frac{C_2^2~\tB_{add}^2}{\Big( \frac{\p \rho}{\p \mu} \Big)_{ \mid_{T, \mu_d}}}.
\een
The terms in question are only dissipated by the charge dissipative terms.

\section{Magneto-transport  of the holographic system}
In this section we shall elaborate the holographic model of the system, 
i.e., Dirac semimetals with $\zz$ topological charge and chiral anomaly. Our goal is to use holographic model 
of the system in question and calculate its thermodynamic properties, in particular the relations among 
the charge densities and chemical potentials. We shall restrict our attention to the analysis of the simplest
cases as given by the equations (\ref{sf-add})-(\ref{sbf-add}). It has to be stressed that hydrodynamic 
approach does not fix the values of the Boltzmann conductivities $\sigma_F$, $\sigma_{F\tB}$, $\sigma_{BF}$ 
and $\sigma_B$ in the above expressions. It provides the general  conditions \cite{rog18b} 
stemming from the positivity of the entropy production. However, hydrodynamics completely determines the anomaly
related kinetic coefficients.

In the probe limit, we are interested in, one considers the AdS Schwarzschild
black brane background, in five-dimensional Einstein-Chern Simons gravity with 
negative cosmological constant with two $U(1)$-gauge fields. The gauge Chern-Simons terms in gravitational
action constitute the possible interactions of the aforementioned fields, which play the crucial role from the point of view of the chiral anomaly and $\zz$ 
topological charge in the studied system. 
The validity of the probe limit approximation requires $B/T^2 \ll 1,~\tB_{add}/T^2 \ll 1$.

The bulk action provided by the holographic model is composed of terms responsible for chiral anomaly and $\zz$ topological charge and reads 
\ben \label{holact}
S &=& \int dx^5 \sqrt{-g} \Bigg( R + \frac{12}{L^2} - \frac{1}{4}F_{\mu \nu} F^{\mu \nu} - \frac{1}{4}B_{\mu \nu} B^{\mu \nu} 
 + \frac{\alpha_1}{3}\ep^{\mu \nu \rho \delta \tau} A_\mu ~F_{\nu \rho}~F_{\delta \tau}\\ \nonumber
&+& \frac{\alpha_2}{3}\ep^{\mu \nu \rho \delta \tau} B_\mu ~B_{\nu \rho}~B_{\delta \tau}  + \frac{\alpha_3}{3}\ep^{\mu \nu \rho \delta \tau} A_\mu ~F_{\nu \rho}~B_{\delta \tau} 
+ \frac{\alpha_4}{3}\ep^{\mu \nu \rho \delta \tau} B_\mu ~F_{\nu \rho}~B_{\delta \tau} \Bigg).
\een
In what follows, our convention is $\ep^{txyzr} =1$.
The equation of motion for the $U(1)$-gauge fields can be written as
\be
\na_\alpha F^{\alpha \beta} + \alpha_1~\ep^{\beta \mu \nu \rho \delta} F_{\mu \nu} F_{\rho \delta} + \frac{2}{3}\alpha_3~ \ep^{\beta \mu \nu  \rho \delta} F_{\mu \nu} B_{\rho \delta}
+ \frac{\alpha_4}{3} ~\ep^{\beta \mu \nu \rho \delta} B_{\mu \nu} B_{\rho \delta} = 0,
\label{eqmf}
\ee
and for the auxiliary $B_{\mu \nu}$ field it has the form as
\be
\na_\alpha B^{\alpha \beta} + \alpha_2~\ep^{\beta \mu \nu \rho \delta} B_{\mu \nu} B_{\rho \delta} + \frac{2}{3}\alpha_4 ~\ep^{\beta \mu \nu  \rho \delta} B_{\mu \nu} F_{\rho \delta}
+ \frac{\alpha_3}{3} ~\ep^{\beta \mu \nu \rho \delta} F_{\mu \nu} F_{\rho \delta} = 0,
\label{eqmb}
\ee
The components of Maxwell and the additional gauge field are provided by
\be
A_\mu = (\phi(r),~0,~B x,~A_z(r),~0), \qquad B_\mu = (\psi(r),~0,~\tB_{add} x,~B_z(r),~0),
\ee
In our consideration, as the background metric we take the line element of AdS-Schwarzschild five-dimensional black brane
\be
ds^2 = r^2 \Big( - f(r) dt^2 + dx^2 + dy^2 + dz^2 \Big) + \frac{dr^2}{r^2~f(r)},
\ee
where $f(r) = 1 - \frac{r_0^4}{r^4}$ and $r_0$ is the radius of the event horizon. The energy density, entropy density and 
the Hawking temperature for the black brane are given, respectively by
\be
\ep= 3~r_0^4, \qquad s = 4 \pi ~r_0^3, \qquad T=\frac{r_0}{\pi}.
\ee
The equations of motion for the gauge fields in the aforementioned background yield
\ben
\phi''(r)&+& \frac{3}{r} \phi'(r) + \frac{8 \alpha_1}{r^3} B A_z'(r) + \frac{8 \alpha_3}{3 r^3} \Big( A_z'(r) \tB_{add} + B_z'(r) B \Big) 
+ \frac{8 \alpha_4}{3 r^3} B_z'(r) \tB_{add} = 0,\\
\psi''(r) &+& \frac{3}{r} \psi'(r) + \frac{8 \alpha_2}{r^3} \tB_{add} B_z'(r) + \frac{8 \alpha_4}{3 r^3} \Big( B_z'(r) B + A_z'(r) \tB_{add} \Big) +
\frac{8 \alpha_4}{3 r^3} A_z'(r) B = 0,\\ \nonumber
A_z''(r) &+& A_z'(r) \Big(\frac{3}{r} + \frac{f'(r)}{f(r)} \Big) + \frac{8 \alpha_1}{r^3 f(r)}B \phi'(r) + 
\frac{8 \alpha_3}{3 r^3 f(r)}  \Big( \phi'(r) \tB_{add} + \psi'(r) B \Big) \\ 
&+& \frac{8 \alpha_4}{3 r^3 f(r)} \psi'(r) \tB_{add} = 0,\\ \nonumber
B_z''(r) &+& B_z'(r) \Big(\frac{3}{r} + \frac{f'(r)}{f(r)} \Big) + \frac{8 \alpha_2}{r^3 f(r)}\tB_{add} \psi'(r) + 
\frac{8 \alpha_4}{3 r^3 f(r)}  \Big( \psi'(r) B + \phi'(r) \tB_{add} \Big) \\ 
&+& \frac{8 \alpha_3}{3 r^3 f(r)} \phi'(r) B = 0,
\een
where the prime denotes the derivative with respect to r-coordinate.

The above set of differential equation can be simplified to the following forms:
\ben
\Big[ r^3~\phi'(r)  &+&  A_z(r)~a_1 + B_z~c \Big]' = 0,\\
\Big[ r^3~f(r)~A_z'(r) &+& \phi(r)~a_1 + \psi(r)~c \Big]' =0,\\ 
\Big[ r^3~\psi'(r)  &+&  B_z(r)~b_1 + A_z~c \Big]' = 0,\\
\Big[ r^3~f(r)~B_z'(r) &+& \psi(r)~b_1 + \phi(r)~c \Big]' =0,
\een
where for the brevity of the notation we have introduced the quantities defined as follows:
\ben
a_1 &=& 8 \Big( \alpha_1~B + \frac{\alpha_3}{3}~\tB_{add} \Big),\\
b_1 &=& 8 \Big( \alpha_2~\tB_{add} + \frac{\alpha_4}{3}~B \Big),\\
c &=& \frac{8}{3} \Big(  \alpha_4 \tB_{add} + \alpha_3 B \Big).
\een
Solving the above set of equations for $\phi$ and $\psi$, with the boundary conditions that at the event horizon of the considered black brane 
one has $\phi(r_0) = \psi(r_0) = 0$, after changing the coordinates given by the relation
\be
u = \frac{r_0^2}{r^2},
\ee
one arrives at the following set of the differential equations:
\ben
\frac{d^2 \phi(u)}{d u^2} &=& \frac{1}{4 r_0^4 (1-u^2)} \Big(a_1~b_1 -c^2 \Big)~\frac{a_1}{b_1}~\phi(u),\\
\frac{d^2 \psi(u)}{d u^2} &=& \frac{1}{4 r_0^4 (1-u^2)} \Big(a_1~b_1 -c^2 \Big)~\psi(u).
\een
The analytical solution near the black brane event horizon may be written as
\ben
\phi_i (u) &=& G_1~{}_2 F_1 \Big[ \frac{1}{4}\Big( - \sqrt{1 - K_i}-1 \Big);~\frac{1}{4}\Big( \sqrt{1 - K_i}-1 \Big);~\frac{1}{2};~u^2 \Big]\\
&+& G_2~u~{}_2 F_1 \Big[ \frac{1}{4}\Big( 1-\sqrt{1 - K_i}\Big);~\frac{1}{4}\Big( 1+ \sqrt{1 - K_i}\Big);~\frac{1}{2};~u^2 \Big],
\een
where $i=\phi,~\psi$ and $K_i$ are given by
\be
K_\psi = \frac{(a_1 b_1 - c^2)}{r_0^4}, \qquad
K_\phi = K_\psi~ \frac{a_1}{b_1}.
\ee
$G_1$ and $G_2$ stand for constants.
Having in mind that near $u \rightarrow 0$ the solutions for $\phi(r)$ and $\psi(r)$ behave like
\be
\phi(r) = \mu - \frac{\rho}{2 r_0^2}~u + \dots, \qquad \psi(r) = \mu_d - \frac{\rho_d}{2 r_0^2} ~u + \dots,
\ee
one obtains the dual charge density for the Maxwell field
\be
\rho = 4~\mu~ r_0^2 ~\frac{\Gamma\Big[\frac{5 - \sqrt{1 -K_\phi}}{4}\Big] ~\Gamma\Big[\frac{5 + \sqrt{1 -K_\phi}}{4}\Big] }
{\Gamma\Big[\frac{3 - \sqrt{1 -K_\phi}}{4}\Big] ~\Gamma\Big[\frac{3 + \sqrt{1 -K_\phi}}{4}\Big] }.
\label{dens}
\ee
For the additional $U(1)$-gauge field, it implies
\be
\rho_d= 4~\mu_d~ r_0^2 ~\frac{\Gamma\Big[\frac{5 - \sqrt{1 -K_\psi}}{4}\Big] ~\Gamma\Big[\frac{5 + \sqrt{1 -K_\psi}}{4}\Big] }
{\Gamma\Big[\frac{3 - \sqrt{1 -K_\psi}}{4}\Big] ~\Gamma\Big[\frac{3 + \sqrt{1 -K_\psi}}{4}\Big] }.
\ee
Let us now apply the holographic description for the system in zero density limits, i.e.,
$\rho = \rho_d \rightarrow 0$,~$\omega_c = \omega_{cd} \rightarrow \omega$ and the additional requirement of vanishing the adequate magnetic field.
For example, by virtue of the equations (\ref{sf-add}) and (\ref{sbf-add})
we can find the explicit forms of the conductivities, when $\tB_{add} =0$. They are provided by
\ben
{\tilde \sigma}_F &=& \sigma_F + \frac{i}{\omega} \frac{C_1^2~B^2}{4~\pi^2~T^2}~\frac{\Gamma\Big[\frac{3 - \sqrt{1 -K_\phi}}{4}\Big] ~\Gamma\Big[\frac{3 + \sqrt{1 -K_\phi}}{4}\Big] }
{\Gamma\Big[\frac{5 - \sqrt{1 -K_\phi}}{4}\Big] ~\Gamma\Big[\frac{5 + \sqrt{1 -K_\phi}}{4}\Big] },\\
{\tilde \sigma}_{B} &=& \sigma_B + \frac{i}{\omega} \frac{C_3^2~B^2}{4~\pi^2~T^2}~\frac{\Gamma\Big[\frac{3 - \sqrt{1 -K_\psi}}{4}\Big] ~\Gamma\Big[\frac{3 + \sqrt{1 -K_\psi}}{4}\Big] }
{\Gamma\Big[\frac{5 - \sqrt{1 -K_\psi}}{4}\Big] ~\Gamma\Big[\frac{5 + \sqrt{1 -K_\psi}}{4}\Big] }.
\een
Similar calculations can be conducted for the case when $B=0$, described by the equations (\ref{sdf}) and (\ref{sdbf}).

In the hydrodynamical limit, we assume that $B \ll T^2$ and $\tB_{add} \ll T^2$.
It implies that the quantities $K_\phi \ll 1$  and $K_\psi \ll 1$. Thus, in the limit under inspection, we receive that the dual densities are provided by
 \be
\rho = 2~\mu~ r_0^2 ~\Big( 1 + \cO \big( \frac{B^2}{T^4},~\frac{\tB_{add}^2}{T^4} \big) \Big), \qquad
\rho_d = 2~\mu_d~ r_0^2 ~\Big( 1 + \cO \big( \frac{B^2}{T^2},   ~\frac{\tB_{add}^2}{T^4}    \big) \Big).
\ee
Consequently, the adequate conductivities are given by expressions
\ben
{\tilde \sigma}_F &=& \sigma_F + \frac{i}{\omega} \frac{C_1^2~B^2}{4~\pi^2~T^2} + \cO \big( \frac{B^2}{T^2},   ~\frac{\tB_{add}^2}{T^4}    \big),\\
{\tilde \sigma}_{B} &=& \sigma_B + \frac{i}{\omega} \frac{C_3^2~B^2}{4~\pi^2~T^2} + \cO \big( \frac{B^2}{T^2},   ~\frac{\tB_{add}^2}{T^4}    \big).
\een
It can be remarked that in the case when $\tB_{add} = 0$ and respectively $C_3= 0$, we arrive at the results presented in \cite{lan15}, where
the only one gauge field was considered and anomaly was bounded with electric and magnetic components of the ordinary Maxwell field.

\begin{figure}
\centerline{\includegraphics[width=0.65\linewidth]{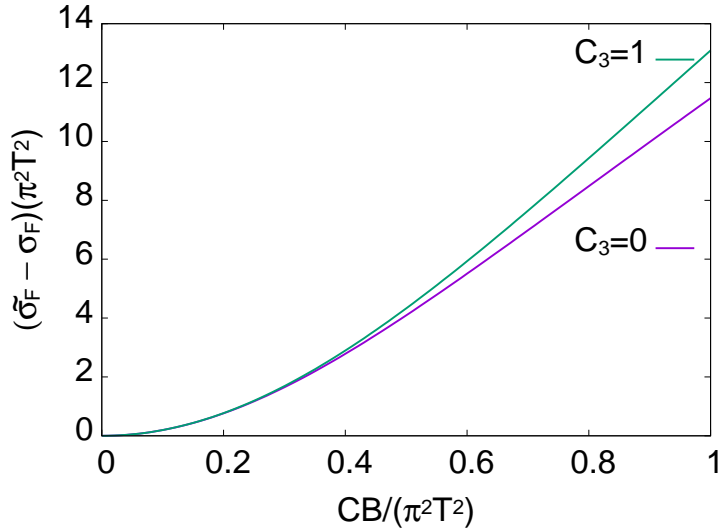}}
\caption{The magnetic field dependence of the real part of 
conductivity $\tilde{\sigma}_F$ calculated from the equation (\ref{sf-add}) for
a system described by (\ref{dens}).  
The upper curve is obtained for all $C_i\neq 0$, while the lower one
for $C_1=C=1$ and $C_3=0$. The existence of $\zz$ topological charge leads to
narrowing of the magneto-conductivity. For actual calculations we 
have used $T=1$, $\tau_e=0.1$, $\omega =0$.}
\label{rys1}
\end{figure}
In  figure \ref{rys1} we plot the magnetic field dependence 
of the relative conductivity $(\tilde{\sigma}_F-\sigma_F)/(\pi^2T^2)$ for a system with the density 
given by the formula (\ref{dens}) and assuming that $B_{add}$ vanishes.
This assumption is valid for materials like Na$_3$Bi (or Cd$_2$As$_3$), 
 in which the spin projection is related to
$\zz$ topological charge. In such systems one does not expect spin analog of the magnetic field
and this justifies our assumption.
Two curves in the figure correspond to the systems with and without $\zz$ topological charge.
The lower curve, calculated for $C_1=C=1, C_3=0=C_2=C_4$ describes the system with chiral anomaly
but without $\zz$ topological charge.
The modifications due to the $\zz$ charge are depicted in the upper curve.
It occurs that the presence of the $\zz$ topological charge leads to the narrowing of the
magneto-conductivity curve. This conclusion seem to agree with the one based 
on the kinetic equation approach to  Na$_3$Bi material \cite{burkov2016}.

The plot has been obtained for  $T=1$, $\tau_e=0.1$, and $\omega =0$.
  In the next section, we calculate the
conductivity $\tilde{\sigma}_F$ using the holographic approach to the problem
and we shall obtain $\sigma_F$ 
part  of the conductivity, as well as,  its magnetic field dependence. 
The effect of $\zz$ topological charge shows up as a narrowing of the magneto-conductivity line.
However, the overall dependence on the $CB/(\pi^2T^2)$ for the system with 
additional topological charge is similar to that with the chiral symmetry only.

\subsection{Relation between hydrodynamic and holographic parameters }
In order to connect the hydrodynamical and holographic descriptions we shall find the 
relation between  $C_i$ and $\alpha_i$ constants.
In order to obtain them we expand the action (\ref{holact}), to the second order in perturbations of both gauge  fields and the metric
\be
A_\mu \rightarrow A_\mu + a_\mu, \qquad B_\mu \rightarrow B_\mu + b_\mu, \qquad g_{\mu \nu} \rightarrow g_{\mu \nu} + h_{\mu \nu}.
\ee
We focus on the part of the action responsible to the first order perturbations connected with currents of gauge fields.
Namely, to the first order in gauge field fluctuations, the adequate part of the action reduces to a boundary term of the following form:
\ben \nonumber
\delta S^{(1)} &=& \int d^4x~\Bigg[ \sqrt{-g} \Big(
F^{\beta r} + \frac{4 \alpha_1}{3} \ep^{\beta \mu \nu \rho} A_\mu F_{\nu \rho} + \frac{2 \alpha_3}{3} \ep^{\beta \mu \nu \rho} A_\mu B_{\nu \rho}
+ \frac{2 \alpha_4}{3} \ep^{\beta \mu \nu \rho} B_\mu B_{\nu \rho}\Big) a_\beta \\
&+& 
\sqrt{-g} \Big(
B^{\beta r} + \frac{4 \alpha_2}{3} \ep^{\beta \mu \nu \rho} B_\mu B_{\nu \rho} + \frac{2 \alpha_3}{3} \ep^{\beta \mu \nu \rho} A_\mu F_{\nu \rho}
+ \frac{2 \alpha_4}{3} \ep^{\beta \mu \nu \rho} B_\mu F_{\nu \rho}\Big) b_\beta \Bigg]_{\mid_{r \rightarrow \infty}}.
\een
Having in mind the above relation, we can read off the forms of the boundary currents
\be
J^\beta_{(F)} = \frac{\delta S^{(1)} }{\delta A_\mu}_{\mid_{r \rightarrow \infty}}, \qquad
J^\beta_{(B)} = \frac{\delta S^{(1)} }{\delta B_\mu}_{\mid_{r \rightarrow \infty}}.
\ee
Using the equations of motion (\ref{eqmf}) and (\ref{eqmb})
and the divergence of $
\na_\beta \Big( J^\beta_{(F)} + J^\beta_{(B)} \Big)$,
we arrive at the relation
\be
C_i = \frac{8~\alpha_i}{3},
\ee
where $i = 1,\dots,4.$

\section{Holographic calculation of conductivities  in the probe limit}
This section is devoted to the direct calculations of the DC-conductivities for the studied system. As in the later section we shall elaborate the probe limit of the holographic model, starting 
from small perturbations in the AdS-Schwarzschild background and computing the longitudinal conductivities from the perturbations.

In order to find the longitudinal holographic conductivity in the model in question, we assume that the fluctuations 
of the vector potentials for both gauge fields
are provided by
\ben
A_\mu &=& (\delta \phi(r) e^{-i \omega t},~0,~0,~B~ x,~\delta A_z (r) e^{- i \omega t} ), \\
B_\mu &=& (\delta \psi(r) e^{-i \omega t},~0,~0,~\tB_{add} ~x,~\delta B_z (r) e^{- i \omega t} ).
\een
The equations of motion for the above perturbations imply
\ben
\delta \phi'(r) &+& \frac{\delta A_z}{r^3}~a_1 + \frac{\delta B_z(r)}{r^3}~c = 0,\\
\delta \psi'(r) &+& \frac{\delta B_z}{r^3}~b_1 + \frac{\delta A_z(r)}{r^3}~c = 0,\\
\delta A_z''(r) &+& \bigg( \frac{3}{r} + \frac{f'(r)}{f(r)} \bigg)~\delta A_z'(r) + \frac{\omega^2}{r^4~f(r)^2} \delta A_z(r) +\frac{\delta \phi'(r)}{r^3~f(r)}~a_1
+ \frac{\delta \psi'(r)}{r^3~f(r)}~c = 0,\\
\delta B_z''(r) &+& \bigg( \frac{3}{r} + \frac{f'(r)}{f(r)} \bigg)~\delta B_z'(r) + \frac{\omega^2}{r^4~f(r)^2} \delta B_z(r) +\frac{\delta \phi'(r)}{r^3~f(r)}~c
+ \frac{\delta \psi'(r)}{r^3~f(r)}~b_1 = 0.
\een
By virtue of the above, one has the following relations for $\delta A_z(r)$ and $\delta B_z(r)$
\ben \label{az}
\delta A_z''(r) &+& \bigg( \frac{3}{r} + \frac{f'(r)}{f(r)} \bigg)~\delta A_z'(r) + \bigg( \frac{\omega^2}{r^4~f(r)^2}  - \frac{\tA}{r^6~f(r)} \bigg)~\delta A_z(r) = 0,\\ \label{bz}
\delta B_z''(r) &+& \bigg( \frac{3}{r} + \frac{f'(r)}{f(r)} \bigg)~\delta B_z'(r) + \bigg( \frac{\omega^2}{r^4~f(r)^2}  - \frac{\tB}{r^6~f(r)} \bigg)~\delta B_z(r) = 0,
\een
where we have introduce the following abbreviations 
\be
\tA = a_1~(a_1 + c) + c~(b_1 + c), \qquad \tB = b_1~(b_1 +c ) + c~ (a_1 + c).
\ee
In $u = r_0^2/r^2$ coordinate the relations (\ref{az})-(\ref{bz}) reduce to the forms
\ben
\delta A_z''(u) &-& \frac{2 u}{1 - u^2}~\delta A_z'(r) + \bigg( \frac{\omega^2}{4~r_0^2~u(1 -u^2)^2}  - \frac{\tA}{4~r_0^4~(1-u^2)} \bigg)~\delta A_z(r) = 0,\\
\delta B_z''(u) &-& \frac{2 u}{1 - u^2}~\delta B_z'(r) + \bigg( \frac{\omega^2}{4~r_0^2~u(1 -u^2)^2}  - \frac{\tB}{4~r_0^4~(1-u^2)} \bigg)~\delta B_z(r) = 0,
\een
where now the prime denotes derivatives with respect to $u$-coordinate.

For the the near horizon limit, i.e.  $u \rightarrow 1$, the above equations can be rewritten as
\ben
\delta A_z''(u) &-& \frac{1}{1 - u}~\delta A_z'(r) + \bigg( \frac{\omega^2}{16~r_0^2~u(1 -u)^2}  - \frac{\tA}{8~r_0^4~(1-u)} \bigg)~\delta A_z(r) = 0,\\
\delta B_z''(u) &-& \frac{1}{1 - u}~\delta B_z'(r) + \bigg( \frac{\omega^2}{16~r_0^2~u(1 -u)^2}  - \frac{\tB}{8~r_0^4~(1-u)} \bigg)~\delta B_z(r) = 0,
\een
with the solution given in terms of the modified Bessel functions
\ben
\delta A_z(u) &=& E_1 (-1)^{- \frac{i \omega}{4 r_0}} I_{- \frac{i \omega}{2r_0}} \bigg[ \sqrt{\frac{\tA}{2}} ~\frac{(1-u)^{\frac{1}{2}}}{r_0^2} \bigg] +
E_2(-1)^{ \frac{i \omega}{4 r_0}} I_{ \frac{i \omega}{2r_0}} \bigg[ \sqrt{\frac{\tA}{2}} ~\frac{(1-u)^{\frac{1}{2}}}{r_0^2} \bigg], \\
\delta B_z(u) &=& D_1(-1)^{- \frac{i \omega}{4 r_0}} I_{- \frac{i \omega}{2r_0}} \bigg[ \sqrt{\frac{\tB}{2}} ~\frac{(1-u)^{\frac{1}{2}}}{r_0^2} \bigg] +
D_2(-1)^{ \frac{i \omega}{4 r_0}} I_{ \frac{i \omega}{2r_0}} \bigg[ \sqrt{\frac{\tB}{2}} ~\frac{(1-u)^{\frac{1}{2}}}{r_0^2} \bigg],
\een
where $E_i,~D_i$ are integration constants. The in-falling boundary conditions for $u$-coordinate correspond to disappearing of $E_2$ and $D_2$.\\
In the far away region $1-u \gg \omega/r_0$, the above relations take forms as
\ben
\delta A_z''(u) &-& \frac{2u}{1 - u^2}~\delta A_z'(r)  - \frac{\tA}{4~r_0^4~(1-u^2)} ~\delta A_z(r) = 0,\\
\delta B_z''(u) &-& \frac{2u}{1 - u^2}~\delta B_z'(r) - \frac{\tB}{4~r_0^4~(1-u^2)} ~\delta B_z(r) = 0,
\een
with the solution in terms of the Legendre functions
\ben
\delta A_z(u) &=& \tE_1 P_{\frac{1}{2} \Big[ \sqrt{1 - \frac{\tA}{r_0^4}} -1 \Big]} (u) + \tE_2 Q_{\frac{1}{2} \Big[ \sqrt{1 - \frac{\tA}{r_0^4}} -1 \Big]} (u),\\
\delta B_z(u) &=& \tD_1 P_{\frac{1}{2} \Big[ \sqrt{1 - \frac{\tB}{r_0^4}} -1 \Big]} (u) + \tD_2 Q_{\frac{1}{2} \Big[ \sqrt{1 - \frac{\tB}{r_0^4}} -1 \Big]} (u),
\een
where $\tE_i,~\tD_i$ are constants.

In order to find the integration constants one should match the near-horizon solution with the far away one, in some intermediate, matching, region 
$\omega/r_0 \ll 1-u \ll 1$. It can be done exactly as in \cite{lan15}. Using the same reasoning it can be revealed that the integration constants fulfill
\ben
\frac{\tE_1}{E_1} &=& 1 + \cO(\omega), \qquad \frac{\tD_1}{D_1} = 1 + \cO(\omega), \\
\frac{\tE_2}{E_1} &=& \frac{i \omega}{2r_0}~(1+ \cO(\omega)), \qquad \frac{\tD_2}{D_1} = \frac{i \omega}{2r_0}~(1+ \cO(\omega)).
\een

To proceed further let us find the value of the Legendre functions at the boundary $u \rightarrow 0$. Namely, one has
\ben 
P_{\frac{1}{2} \Big[ \sqrt{1 - \frac{\tA}{r_0^4}} -1 \Big]} (u \rightarrow 0)
&=& p_1(\tA_i) + p_2(\tA_i) u + \cO(u^2),\\
Q_{\frac{1}{2} \Big[ \sqrt{1 - \frac{\tA}{r_0^4}} -1 \Big]} (u \rightarrow 0) &=&  q_1(\tA_i) + q_2(\tA_i) u + \cO(u^2),
\een
where we have defined
\ben
p_1(\tA_i) &=& \frac{\sqrt{\pi}}{\Gamma\Big[\frac{3 - \sqrt{1 -\frac{\tA}{r_0^4}}}{4}\Big] 
\Gamma\Big[\frac{3 + \sqrt{1 - \frac{\tA}{r_0^4}}}{4}\Big] },  \quad
p_2(\tA_i) = - \frac{\sqrt{\pi}~\tA_i}{8 r_0^4~\Gamma\Big[\frac{5 - \sqrt{1 -\frac{\tA}{r_0^4}}}{4}\Big] \Gamma\Big[\frac{5 + \sqrt{1 -\frac{\tA}{r_0^4}}}{4}\Big] } ,\\
q_1(\tA_i) &=& - \frac{\sqrt{\pi} \sin \Big( \frac{\pi}{4}(\sqrt{1 -\frac{\tA_i}{r_0^4}} -1)\Big) \Gamma\Big[\frac{1+ \sqrt{1 -\frac{\tA}{r_0^4}}}{4}\Big] }
{2 \Gamma\Big[\frac{3 + \sqrt{1 -\frac{\tA}{r_0^4}}}{4}\Big]} , \\
q_2(\tA_i) &=&
\frac{\sqrt{\pi} \cos \Big( \frac{\pi}{4}(\sqrt{1 -\frac{\tA_i}{r_0^4}} -1)\Big) \Gamma\Big[\frac{3+ \sqrt{1 -\frac{\tA}{r_0^4}}}{4}\Big] }
{\Gamma\Big[\frac{1 + \sqrt{1 -\frac{\tA}{r_0^4}}}{4}\Big]}. 
\een
In order to find the conductivities we derive the relations which envisages the ingoing boundary conditions for 
the considered longitudinal currents
\ben
\delta A_z(u) &=& \tE_1~ p_1(F) + \tE_2~ q_1(F) + u (\tE_1~ p_2(F) + \tE_2~ q_2(F)),\\
\delta B_z(u) &=& \tD_1~ p_1(B) + \tD_2~ q_1(B) + u (\tD_1~ p_2(B) + \tD_2~ q_2(B)).
\een

Using the definition of the conductivity \cite{hor08}, where the logarithmic divergence is removed with the adequate boundary counter-term in the gravity action \cite{tay00}, one attains
\be
\sigma = \frac{2~ A^{(0)}_\mu}{i~ \omega ~A^{(2)}_\mu }+ \frac{i~ \omega}{2}, 
\label{sss}
\ee
where the gauge field fall off implies the following:
\be
A_\mu = A^{(0)}_\mu + \frac{A^{(2)}_\mu}{r^2} + \dots,
\label{aaa}
\ee
( the similar condition holds for $B_\mu$ field). Consequently,
by virtue of the equations (\ref{sss}) and (\ref{aaa}), we obtain the expressions describing ${\tilde \sigma}_{F}$ and ${\tilde \sigma}_{B} $. They can be written as
\ben
{\tilde \sigma}_{F} &=& \frac{2r_0^2~(\tE_1~p_2 + \tE_2~q_2)}{i \omega~(\tE_1~p_1 + \tE_2~q_1)} + \frac{i \omega}{2},\\
{\tilde \sigma}_{B} &=& \frac{2r_0^2~(\tD_1~p_2 + \tD_2~q_2)}{i \omega~(\tD_1~p_1 + \tD_2~q_1)} + \frac{i \omega}{2}.
\een
Expansions of ${\tilde \sigma}_{F} $ and ${\tilde \sigma}_{B}$, at the leading order in $\omega$, reveal the relations
\ben \label{eq:433}
{\tilde \sigma}_F &=& \bigg[ \frac{i}{\omega} \frac{\tA}{4 r_0^2}+ \frac{\pi~\tA}{16 r_0^3~\cos \Big( \frac{\pi}{2} \sqrt{1 - \frac{\tA}{r_0^4}}
\Big)}\bigg]
\frac{\Gamma\Big[\frac{3 - \sqrt{1 -\frac{\tA}{r_0^4}}}{4}\Big] 
~\Gamma\Big[\frac{3 + \sqrt{1 - \frac{\tA}{r_0^4}}}{4}\Big] }
{\Gamma\Big[\frac{5 - \sqrt{1 -\frac{\tA}{r_0^4}}}{4}\Big] ~\Gamma\Big[\frac{5 + \sqrt{1 -\frac{\tA}{r_0^4}}}{4}\Big] }
,\\
{\tilde \sigma}_{B} &=& \bigg[ \frac{i}{\omega} \frac{\tB}{4 r_0^2}+ \frac{\pi~\tB}{16 r_0^3~\cos \Big( \frac{\pi}{2} \sqrt{1 - \frac{\tB}{r_0^4}}
\Big)}\bigg]
\frac{\Gamma\bigg[\frac{3 - \sqrt{1 -\frac{\tB}{r_0^4}}}{4}\bigg] 
~\Gamma\bigg[\frac{3 + \sqrt{1 - \frac{\tB}{r_0^4}}}{4}\bigg] }
{\Gamma\bigg[\frac{5 - \sqrt{1 -\frac{\tB}{r_0^4}}}{4}\bigg] ~
\Gamma\bigg[\frac{5 + \sqrt{1 -\frac{\tB}{r_0^4}}}{4}\bigg] },
\een
In the limit when $B/T^2 \ll 1$ and $\tB_{add}/T^2 \ll 1$, which is equivalent to the condition that $\tA \ll 1$ and $\tB \ll 1$, one gains the formulae  
\ben
{\tilde \sigma}_{F} &=& \sigma_F + \frac{i}{\omega}~\frac{\tA}{2 \pi^2 T^2} + \cO \Big(\frac{\tA^2}{T^4} \Big), \qquad
\sigma_F = \frac{\pi ~T}{2} + \cO \Big(\frac{\tA^2}{T^4} \Big),\\
{\tilde \sigma}_{B} &=& \sigma_B + \frac{i}{\omega}~\frac{\tB}{2 \pi^2 T^2} + \cO \Big(\frac{\tB^2}{T^4} \Big), \qquad
\sigma_B = \frac{\pi ~T}{2} + \cO \Big(\frac{\tB^2}{T^4} \Big).
\een
Consequently, in the limit $\tB_{add} \rightarrow 0$, the conductivities imply
\ben
{\tilde \sigma}_{F} &=& \sigma_F + \frac{i}{\omega} \frac{(\talpha ~B)^2}{2 \pi^2 T^2} + \cO \Big( \talpha^4 B^4 \Big), \qquad
\sigma_F = \pi ~T + \cO \Big( \talpha^4 B^4 \Big),\\
{\tilde \sigma}_{B} &=& \sigma_B + \frac{i}{\omega} \frac{({\tilde \talpha} ~B)^2}{2 \pi^2 T^2} + \cO \Big( {\tilde \talpha}^4 B^4 \Big), \qquad
\sigma_B = \pi ~T + \cO \Big( {\tilde \talpha}^4 B^4 \Big),
\een
where we have defined
\ben
\talpha^2 &=& 64 \alpha_1^2 + \frac{64}{9} \big( \alpha_3^2 + \alpha_3 \alpha_4 \big) + \frac{64}{3} \alpha_3 \alpha_1,\\
{\tilde \talpha}^2 &=& \frac{64}{9} \alpha_4^2 + \frac{64}{9} \big( \alpha_3^2 + \alpha_3 \alpha_1 \big) + \frac{64}{3} \alpha_3 \alpha_1.
\een

Turning our attention to the limit when $\tA \gg 1$ and $\tB \gg 1$, one arrives at
\ben
{\tilde \sigma}_{F} &=& \sigma_F + \frac{i}{\omega} 2~\sqrt{\tA} + \cO \Big( \frac{1}{2 \sqrt{\tA}} \Big), \qquad
\sigma_F = \frac{T}{2}~e^{- \frac{\sqrt{\tA}}{2 \pi T^2}}~\bigg( \frac{\sqrt{\tA}}{T^2} + \cO \Big( \frac{1}{2 \sqrt{\tA}} \Big)\bigg),\\
{\tilde \sigma}_{B} &=& \sigma_B  + \frac{i}{\omega} 2~\sqrt{\tB} + \cO \Big( \frac{1}{2 \sqrt{\tB}} \Big), \qquad
\sigma_B = \frac{T}{2}~e^{- \frac{\sqrt{\tB}}{2 \pi T^2}}~\bigg( \frac{\sqrt{\tB}}{T^2} + \cO \Big( \frac{1}{2 \sqrt{\tB}} \Big) \bigg).
\een

\begin{figure}
\begin{center}
\includegraphics[width=0.45\linewidth]{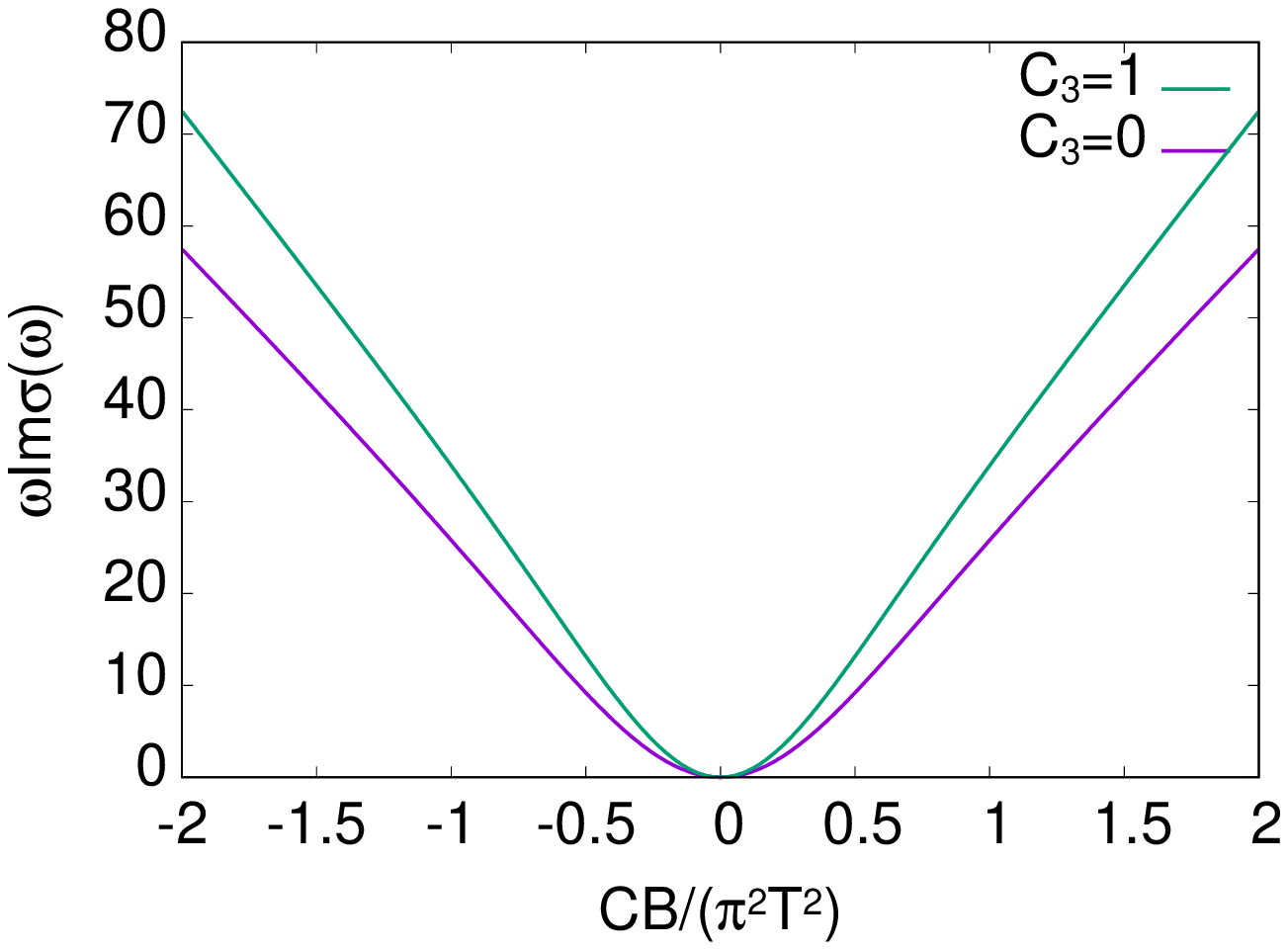}
\includegraphics[width=0.45\linewidth]{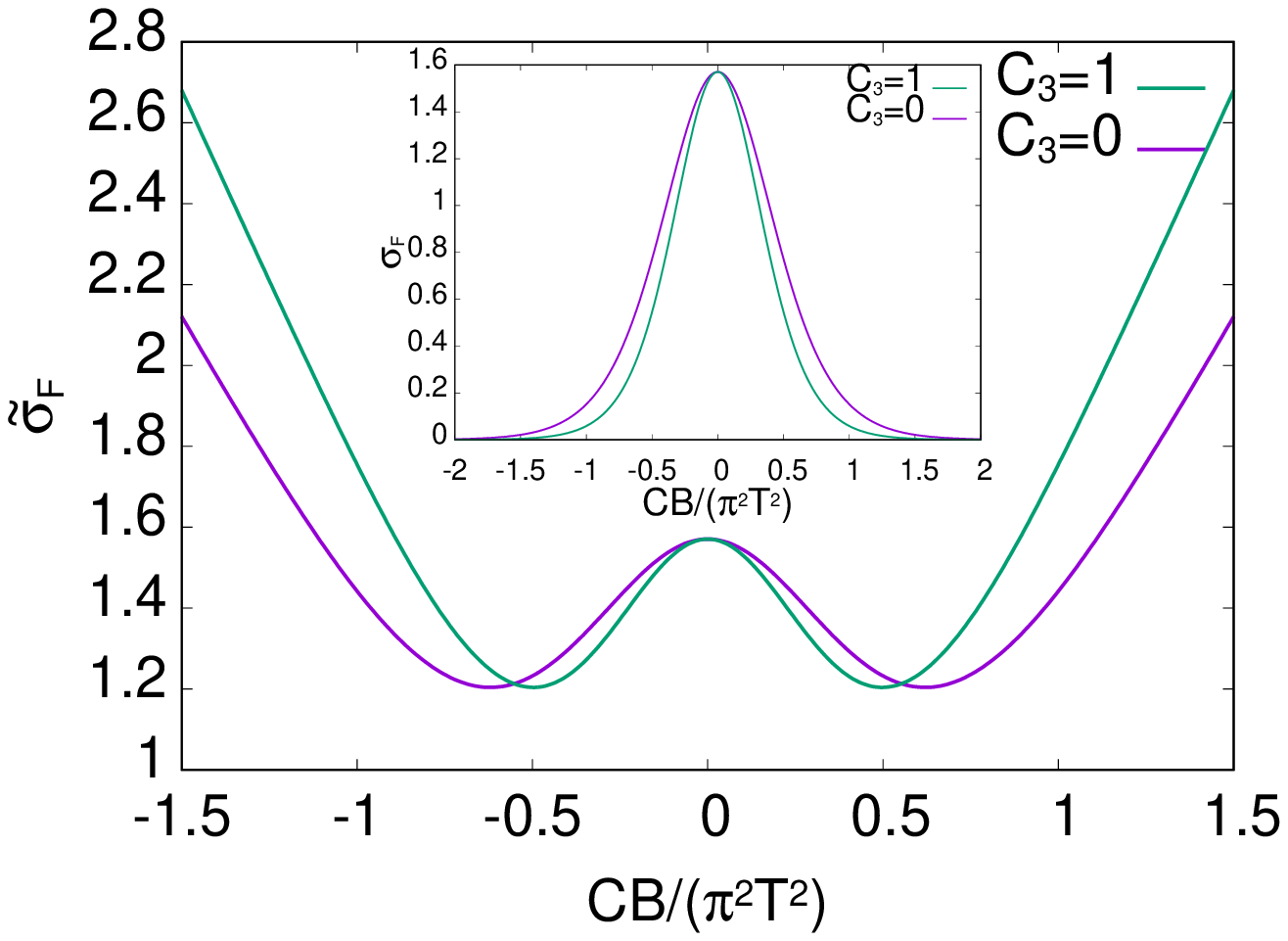}
\includegraphics[width=0.65\linewidth]{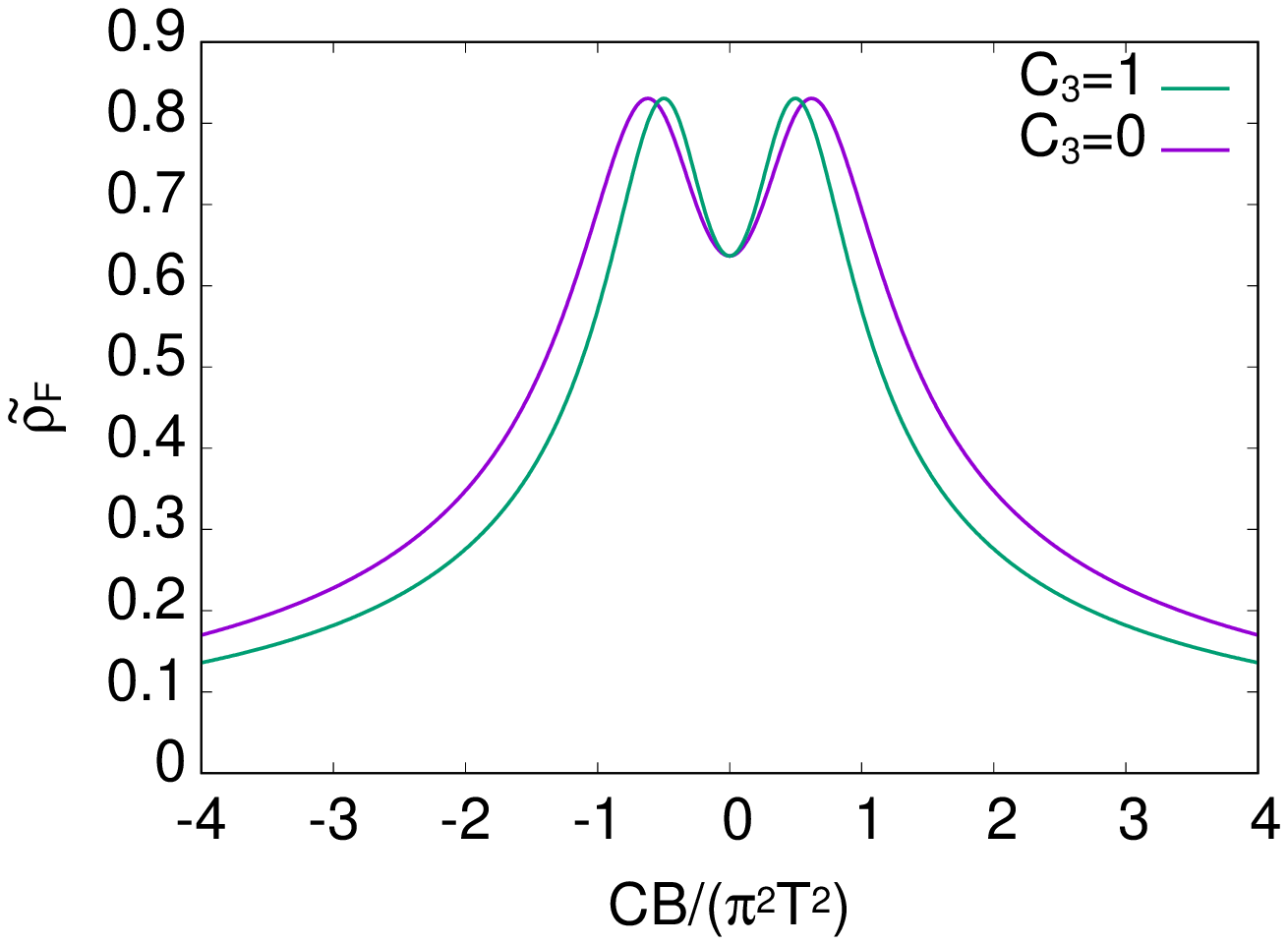}
\end{center}
\caption{The   dependence of the $\omega \textrm{Im}\sigma(\omega)$ (upper left panel)
and the real part of the DC conductivity $\tilde{\sigma}_F(\omega=0)$ (upper right panel) 
on $\frac{CB}{\pi^2T^2}$, 
with $C=C_1$ being a chiral anomaly parameter. The conductivity has been
  calculated from the relation (\ref{eq:433}) for
a holographic system with action (\ref{holact}). We used the  relation 
$C_i=8\alpha_i/3$ between anomaly parameters $C_i$ in the hydrodynamic approach and
corresponding parameters $\alpha_i$ adopted in (\ref{holact}).  
One of the two curves in each panel is obtained for all $C_i\neq 0$ (denoted $C_3=1$), while the other 
for $C_1=C=1$ and $C_3=0$ (denoted $C_3=0$). The existence of $\zz$ topological charge
(corresponding to $C_3=1=C_2=C_4=C_1$ and shown 
by the upper curve in the upper panel)  leads to the narrowing of the magneto-conductivity. 
To obtain the magnetoconductivity shown in the upper right panel 
we have assumed that $T=1$, $\tau_e=0.05$ and $\omega =0$. The inset shows the Boltzmann part $\sigma_F$.
 The lower panel of the figure  shows the magnetoresistance of the same system.}
\label{rys2}
\end{figure}

It will be interesting to analyze the effect of the additional $\zz$ topological charge
on the transport properties of the holographic system under consideration. First we note that the
holographic approach allowed us the direct calculation of $\sigma_F$, the component 
of conductivity undetermined in the hydrodynamic attitude. In the literature it is
sometimes called quantum critical component of conductivity albeit to call it
the Boltzmann conductivity is perhaps more appropriate in the present context, 
as it can be obtained from the Boltzmann kinetic 
equation. One has to remember that the general formula (\ref{eq:433}) is valid to lowest
order in frequency. The conductivity $\sigma_F$ depends on the magnetic field $B$ and also
on the additional magnetic field $B_{add}$. In accordance with the paper \cite{burkov2016},
we assume here $B_{add}=0$ arguing, after the cited paper, that $\zz$ topological charge, at least in some systems,
is related to the spin projections and one does not expect spin analog of the $B$ magnetic field.

In the upper left panel of the figure \ref{rys2},  we show the dependence 
of $\omega \textrm{Im}\sigma(\omega)$
on the magnetic field, and more exactly, on the parameter   $\frac{CB}{\pi^2T^2}$, for the two investigated
cases. One curve ($C_3=0$) corresponds to the system with the chiral anomaly described  by $C_1=C\neq 0$ with $C_2=C_3=C_4=0$. The presence 
of $\zz$ topological charge in the figure represented by the curve with
$C_3=1$~$(=C_2=C_4)$, has an effect 
of narrowing the dependence of $\omega \textrm{Im}\sigma(\omega)$ on the magnetic field. This result harmonizes with the conclusion obtained in the previous section 
by applying the hydrodynamic approach to the holographic system.  

The upper right panel of the figure  \ref{rys2} illustrates similar  dependence of the $\tilde{\sigma}_F$ 
on the  scaled magnetic field for the system with only chiral symmetry ($C_3=0$) and for 
the type II Dirac semimetal with both chiral anomaly and $\zz$ topological charge. 
The real part of the DC-conductivity (i.e., $\omega=0$) 
$\tilde{\sigma}_F$, shown in the main part of the figure, comprises the quantum critical (or Boltzmann) part and the
frequency dependent part which remains nonzero if the system is characterized by the finite charge relaxation 
time $\tau_e=0.05$ introduced phenomenologically by the substitution $\omega \rightarrow \omega +i/\tau_e$. 
 The inset to the figure shows the dependence on $\frac{CB}{\pi^2T^2}$ of the Boltzmann contribution to the total conductivity. Interestingly it features strong non-monotonic 
dependence on the B-field, which is only limited to relatively low fields. 
Summing it all up, we have obtained a non-monotonous overall dependence of the magneto-conductivity on the magnetic field. Again the 
presence of the $\zz$ topological charge reveals a narrowing of the magneto-conductivity line.

The lower panel of the figure \ref{rys2} depicts the  magnetoresistivity 
of the same system. One  observes that for the very small magnetic field
the DC-magnetoresistivity $[\tilde{\rho_F}(B)-\tilde{\rho}_F(0)]/\tilde{\rho}_F(0)$ 
(with $\tilde{\rho}_F=1/\tilde{\sigma}_F$) is positive and only 
for higher fields it becomes negative. The low field negative magneto-conductivity and positive 
magnetoresistivity, followed by the negative magnetoresistivity at high fields, has been 
observed experimentally \cite{li16} in ZrTe$_5$. The holographic attitude provides qualitative  the explanation 
of the experimental findings competitive to the weak anti-localization  proposal.

Besides narrowing of the $\sigma_F(B)$ curve, the auxiliary topological charge, changes the relative 
contribution of quantum critical and relaxation limited parts. This is visible as a shift
of the minimum of the two magneto-conductivity curves. The existence of $\zz$ topological charge shifts that point towards 
lower magnetic fields. However, the detailed behavior also depends on the assumed value
of the charge relaxation time, $\tau_c$.

\section{Summary and conclusions}
\label{sum-concl}
We have studied the behavior of the longitudinal conductivities in the Weyl semimetals with $\zz$ topological charge and chiral anomaly in the hydrodynamics limit,
with the adequate magnetic fields which are connected with the ordinary Maxwell and the second $U(1)$-gauge fields. The auxiliary gauge field was introduced in order
to envisage the $\zz$ anomalous charge.

Using the kinetic coefficients \cite{rog18b} for the system in question, being the functions of
temperature, $\rho,~\rho_d,~\mu,~\mu_d$, we found the DC-conductivities in the presence of magnetic fields. In our considerations we take into account the
dissipation terms, i.e., momentum, energy and charges dissipation terms. For the general case we obtain the complicated functions which depend
 on both magnetic fields, as well as, chiral anomaly and $\zz$ topological charge coefficients. In order to simplify these relations we have considered limiting cases.
 Namely, the cases of $\rho=\mu=0$,~$\rho_d=\mu_d=0$ and $\rho=\mu=\rho_d=\mu_d=0$ were elaborated. Moreover when one assumes that the one of
 the considered magnetic fields is equal to zero, we get the DC-conductivities which are functions of charge densities. Namely, if
 $B=0$, the conductivity depends on $\rho_d$, and on the contrary, when $\tB_{add} =0$, it is the function of $\rho$.

In the case when $\rho=\mu=\rho_d=\mu_d=0$ and vanishing of both magnetic fields, one obtains the result that even at zero densities we have finite
DC-conductivities. The received relations were only dissipated by the charge dissipation terms responsible for both $U(1)$-gauge fields.
In the limit when the additional gauge field is equal to zero, one gets the results presented in \cite{lan15}.

The system was also examined by the holographic attitude, by means of five-dimensional Einstein-Chern-Simons gravity. The gauge Chern-Simons
terms in gravitational action were associated with the possible interactions, playing the crucial role from the point of view of chiral anomaly and $\zz$ topological charge.
As a background, in the probe limit attitude, we take the line element of five-dimensional AdS-Schwarzschild black brane.
Starting from the small perturbations of the aforementioned black brane we have found the longitudinal conductivities for the considered model.
We directly found that using Kubo formula, in the probe limit, the results match the hydrodynamical approach.

In the holographic description we did not include the dissipation terms. In addition, in the case when the additional field is equal to zero, 
we obtained the previously announced results \cite{lan15}.

The future experimental observations may put some restrictions on the constants $C_i$ connected with the magneto-conductivities. Namely,
the measurements similar to those conducted in \cite{goo17} may shed some light on the problem in question. Our results show that the presence of $\zz$
topological charge causes the narrowing of the magneto-conductivity curve. On the other hand,
the temperature dependence and the narrowing of the magneto-conductivity curve 
qualitatively agree with the experimental data \cite{zha15,su2015,neupane2014,neupane2016}.

\acknowledgments
The authors acknowledge the partial support of this work 
by the  National Science Center (Poland) through the grant no. DEC-2014/15/B/ST2/00089.




\end{document}